\documentclass[
reprint,
showpacs,
showkeys,
 amsmath,amssymb,
 aps,
 prc,
floatfix,
]{revtex4-1}

\usepackage[utf8]{inputenc}
\usepackage{graphicx}
\usepackage{color}
\usepackage{enumerate}

\usepackage{qmech-revtex}

\begin{document}

\preprint{}

\title{A coarse-graining approach for dilepton production at SPS energies}

\author{Stephan Endres}
 \email{endres@th.physik.uni-frankfurt.de}
\author{Hendrik van Hees}%
\author{Janus Weil}%
\author{Marcus Bleicher}%
\affiliation{%
Frankfurt Institute for Advanced Studies,
Ruth-Moufang-Strasse 1, D-60438 Frankfurt, Germany
}%
\affiliation{
Institut f{\"u}r Theoretische Physik, Universit{\"a}t Frankfurt,
Max-von-Laue-Strasse 1, D-60438 Frankfurt, Germany
}

\date{\today}

\begin{abstract}
  Coarse-grained output from transport calculations is used to determine
  thermal dilepton emission rates by applying medium-modified spectral
  functions from thermal quantum field theoretical models. By averaging
  over an ensemble of events generated with the UrQMD transport model,
  we extract the local thermodynamic properties at each time step of the
  calculation. With an equation of state the temperature $T$ and
  chemical potential $\mu_{\mathrm{B}}$ can be determined. The approach
  goes beyond simplified fireball models of the bulk-medium evolution by
  treating the full (3+1)-dimensional expansion of the system with
  realistic time and density profiles. For the calculation of thermal
  dilepton rates we use the in-medium spectral function of the $\rho$
  meson developed by Rapp and Wambach and consider thermal QGP and
  multi-pion contributions as well. The approach is applied to the
  evaluation of dimuon production in In+In collisions at top SPS
  energy. Comparison to the experimental results of the NA60 experiment
  shows good agreement of this ansatz. We find that the experimentally
  observed low-mass dilepton excess in the mass region from 0.2 to 0.6
  GeV can be explained by a broadening of the $\rho$ spectral function
  with a small mass shift. In contrast, the intermediate mass region
  ($M > 1.5$ GeV) is dominated by a contribution from the quark-gluon
  plasma. These findings agree with previous calculations with fireball
  parametrizations. This agreement in spite of differences in the
  reaction dynamics between both approaches indicates that the
  time-integrated dilepton spectra are not very sensitive to
  details of the space-time evolution of the collision.
\end{abstract}



\maketitle

\section{\label{sec:Intro} Introduction }

The in-medium properties of hadrons have been a field of intense studies
in theory and experiment over the last years
\cite{Hatsuda:1991ez,Rapp:1999ej,Brown:2001nh,Hayano:2008vn,Rapp:2009yu,Leupold:2009kz}.
One aims to learn more about the phase diagram of QCD and especially to
find hints for a possible restoration of chiral symmetry. For such
studies of hot and dense nuclear matter, dileptons are a unique tool. As
they do not interact strongly, they suffer only negligible final-state
interactions with the medium and thus provide insight into the spectral
properties of their source, i.e., dileptons provide a direct view on the
in-medium electromagnetic current-current correlation function of QCD
matter during the entire history of the collision, from first
nucleon-nucleon reactions to final freeze-out
\cite{Rapp:1999ej,Xia:1988ym,Gale:2003iz}. However, this advantage also
comes with a drawback. As we only get time-integrated spectra over the
whole space-time evolution of a nuclear reaction, it is difficult to 
disentangle the various contributing processes which requires a
realistic description of all collision stages.

Considering the different experimental efforts to investigate dilepton
production in heavy-ion collisions, the NA60 experiment plays a
prominent role. It measured dimuons in heavy-ion collisions at top SPS
energies with an unprecedented precision. The high accuracy of the
measurement enabled the subtraction of the background contributions
(long-lived mesons as $\eta$, $\eta'$, $\omega$, $\phi$) from the
dilepton spectra. Consequently, the NA60 results deliver direct insight
into the in-medium effects on the $\rho$ spectral function in the
low-mass region up to 1\,GeV
\cite{Arnaldi:2006jq,Arnaldi:2007ru,Arnaldi:2008fw} and the thermal
dimuon emission in the intermediate mass region \cite{Arnaldi:2008er}. A
main finding was a large excess of lepton pairs in the mass region
0.2-0.4\,GeV, which confirmed the previous results by CERES
\cite{Agakishiev:1995xb}. Theoretical studies showed that this excess
can be explained by a strong broadening of the $\rho$ spectral function
with small mass shifts
\cite{vanHees:2006ng,vanHees:2007th,Dusling:2006yv,Ruppert:2007cr}.

In general there exist two different types of approaches to describe
heavy-ion collisions, microscopic and macroscopic ones. The
microscopic models, e.g., transport models as UrQMD
\cite{Bass:1998ca,Bleicher:1999xi}, HSD \cite{Ehehalt:1996uq} or GiBUU
\cite{Buss:2011mx}, focus on the description of all the subsequent
hadron-hadron collisions (respectively interactions of partons),
according to the Boltzmann equation. The difficulty here is to implement
in-medium effects in such a microscopic non-equilibrium approach which
is highly non-trivial, but nevertheless some investigations on that
issue have been conducted successfully
\cite{Cassing:1997jz,Schenke:2005ry,Schenke:2006uh,Schenke:2007zz,Bratkovskaya:2008bf,Cassing:2009vt,Barz:2009yz,Linnyk:2011hz,Weil:2012ji,Weil:2012qh}. On
the other hand, in macroscopic models such as thermal fireball models
\cite{Rapp:2013nxa} or hydrodynamics
\cite{Teaney:2001av,Hirano:2002ds,Kolb:2003dz,Nonaka:2006yn,Dusling:2006yv,Vujanovic:2013jpa},
the application of in-medium hadronic spectral functions from thermal
quantum-field theory is straightforward. However, due to their plainness
the fireball parametrizations might oversimplify the real dynamics of a
nuclear reaction, and hydrodynamical simulations may not be applicable
to the less hot and dense medium created at lower collision
energies. Furthermore, the creation of an equilibrium state of hot and
dense matter after quite short formation times is usually assumed in
these models whereas the results from microscopic investigations
indicate the importance of non-equilibrium effects during the evolution
of a heavy-ion collision \cite{Bravina:1999dh}.

Combining a realistic (3+1)-dimensional expansion of the system with
full in-medium spectral functions for the thermal emission of dileptons
is yet an important challenge for theory. One approach that has proven
successful in explaining the NA60 results is the investigation of
dilepton production with a hybrid model \cite{Santini:2011zw}. It
combines a cascade calculation of the reaction dynamics with thermal
emission from an intermediate hydrodynamic stage. However, as all
hydro-approaches it is only working properly for sufficiently large
collision energies. Furthermore, the hybrid-approach falls into three
different stages, a pre-hydro phase, the hydrodynamic stage and the
transport phase after particlization. An application of in-medium
spectral functions hereby only applies for the rather short hydro stage.

For the study presented in this paper we follow an approach which uses a microscopic
description for the whole evolution of the collision and enables the use
of in-medium spectral functions from thermal quantum-field theoretical
models at all stages. Taking a large number of events generated with the
Ultra-relativistic Quantum Molecular Dynamics (UrQMD) model, we place
the output on a space-time grid and extract the local temperature and
baryon chemical potential by averaging energy and baryon density over
the events (i.e., we ``coarse-grain'' the microscopic results) which
allows for the calculation of local thermal dilepton emission. This
ansatz was previously proposed and used to calculate hadron, dilepton
and photon spectra \cite{Huovinen:2002im}. For the present work we
modify the approach to include also the very initial stage of the
reaction (which was separately treated in the cited work) and account
for non-equilibrium effects with respect to the pion
dynamics. Additionally we include non-thermal contributions to really
cover the whole evolution of the nucleus-nucleus reaction.

This paper is structured as follows. In Section \ref{sec:Model} the
coarse-graining approach is described in detail and the different
contributions to the dilepton emission included in the model are
introduced. Subsequently we present the results for the space-time
evolution of the nuclear reaction in Section \ref{ssec:STE} followed by
the dilepton invariant-mass and transverse-momentum spectra, which we
compare to the experimental results in \ref{ssec:Results}. Finally, in
Section \ref{sec:Summary} a summary and an outlook on further studies is
given.

\section{\label{sec:Model} The model }
\subsection{\label{ssec:Coarse} The coarse-graining approach }

The underlying input for our calculations stems from the
Ultra-relativistic Quantum Molecular Dynamics Approach (UrQMD)
\cite{Bass:1998ca, Bleicher:1999xi, Petersen:2008kb, UrQMDweb}. It is a
non-equilibrium transport approach that includes all hadronic resonance
states up to a mass of 2.2\,GeV and constitutes an effective solution of
the relativistic Boltzmann equation. A heavy-ion collision is simulated
such that all hadrons are propagated on classical trajectories in
combination with elastic and inelastic binary scatterings and resonance
decays. At higher energies, string excitation is possible as well. The
model has been checked to describe hadronic observables up to RHIC
energies with good accuracy \cite{Petersen:2008kb}. For the further
investigations we use the UrQMD output in time steps.  This provides
positions, momenta, and energies of all particles and resonances at that
specific moment in time. The size of each time step for the present calculations is chosen
as $\Delta t= 0.2$\,fm/$c$.

In the UrQMD model, the particle distribution function of all hadrons is
given by an ensemble of point particles, which at time $t$ are defined
by their positions $\vec{x}_{h}$ and momenta $\vec{p}_{h}$. Each
particle's contribution to the phase-space density is then defined as
\begin{equation}
\delta^{(3)}(\vec{x}-\vec{x}_{h}(t))\delta^{(3)}(\vec{p}-\vec{p}_{h}(t)).
\end{equation} 
With a sufficiently large number of events the distribution function
$f(\vec{x},\vec{p},t)$ takes a smooth form
\begin{equation}
  f(\vec{x},\vec{p},t)=\left\langle\ \sum_{h}
    \delta^{(3)}(\vec{x}-\vec{x}_{h}(t))\delta^{(3)}(\vec{p}-\vec{p}_{h}(t))\right\rangle. 
\end{equation} 
Hereby, the ensemble average $\left\langle \cdot \right\rangle$ is taken
over simulated events. As the UrQMD model constitutes a non-equilibrium
approach, the equilibrium quantities have to be extracted locally at
each space-time point. In consequence we set up a grid of small
space-time cells with a spatial extension $\Delta x$ of 0.8\,fm and
average the UrQMD output for each cell on that grid. One can then
determine the energy-momentum tensor $T^{\mu\nu}$ and the baryon
four-flow according to the following expressions:
\begin{alignat}{2}
  T^{\mu\nu}&=\int
  \dd^{3}p\frac{p_{\mu}p_{\nu}}{p_{0}}f(\vec{x},\vec{p},t) \nonumber
  \\
  &=\frac{1}{\Delta V}\left\langle \sum\limits_{i=1}^{N_{h} \in \Delta
      V} \frac{p_{\mu}^{i}\cdot p_{\nu}^{i}}{p_{0}^{i}}\right\rangle,
  \\
\label{4}
\begin{split}
  j_{\mu}^{\text{B}} &= \int
  \dd^{3}p\frac{p_{\mu}}{p_{0}} f^{\text{B}}(\vec{x},\vec{p},t) \\
&=\frac{1}{\Delta
    V}\left\langle \sum\limits_{i=1}^{N_{\text{B}/\bar{\text{B}}} \in \Delta
      V}\pm\frac{p_{\mu}^{i}}{p_{0}^{i}}\right\rangle.
\end{split}
\end{alignat}
For the net-baryon flow in (\ref{4}) only the distribution function of
baryons and anti-baryons $f^{\mathrm{B}}(\vec{x},\vec{p},t)$ is considered, excluding all
mesons. Each anti-baryon hereby gives a negative contribution to
$j_{\mu}^{\text{B}}$. On the contrary, the distribution function for all
hadrons in the cell enters in $T^{\mu\nu}$.
  
According to Eckart's definition \cite{Eckart:1940te} the local
rest-frame (LRF) is tied to conserved charges, in our case the net-baryon
number. Consequently one has to perform a Lorentz transformation into
the frame, where $\vec{j}^{\text{B}}=0$. The unit vector in direction of
the baryon flow takes the form
\begin{eqnarray}
u^{\mu}=\frac{j^{\mu}}{(j^{\nu}j_{\nu})^{1/2}}=(\gamma,\gamma \vec{v}),
\end{eqnarray}
where $u_{\mu}u^{\mu}=1$. With this, the rest-frame values for the
baryon and energy density are obtained by
\begin{alignat}{2}
  \rho_{B} &= j_{\mu}u^{\mu}=j^{0}_{\text{LRF}}, \\
  \varepsilon &= u_{\mu}T^{\mu\nu}u_{\nu}=T^{00}_{\text{LRF}}.
\end{alignat}

\subsection{Equilibration and thermal properties of the cells}
For the case of a fully equilibrated ideal fluid the energy-momentum
tensor would be completely diagonal in the local rest-frame. An
important question is to which extent equilibrium is obtained in the
present approach and how one can deal with deviations from it.

In macroscopic descriptions of nuclear reaction dynamics local
equilibrium is usually introduced as an ad-hoc assumption. (In fireball
models, one even assumes a globally equilibrated system). The
equilibration of the system is generally assumed to be very fast (on the
scale of 1-2\,fm/$c$), after an initial phase. In contrast, no such
hypothesis is made within transport approaches, where only the
microscopic interactions between the constituent particles of the medium
are described. Furthermore, these approaches implicitly include effects 
as e.g.~viscosities or heat conduction and do not fully account for 
detailed balance. Consequently, non-equilibrium is the normal case at
any stage of the collision. On the other hand, to calculate thermal
emission rates for dileptons from many-body quantum-field theoretical
models, it is necessary know the temperature and baryochemical potential
which are by definition equilibrium properties. But when extracting
thermodynamic properties from a transport model one has to deal with the
problem that many cells are not found in equilibrium. As the
coarse-graining approach aims to treat the entire space-time evolution
of the collision, this case has to be handled with special care.

Ideally, if complete local equilibrium is achieved, then in each cell
the momentum spectrum and the particle abundances should follow a
Maxwell-J\"uttner-distribution
\begin{equation}
\label{maxjuett}
f_{\mathrm{eq}}(p,m_{i})=\exp\left
  (-\frac{\sqrt{p^{2}+m_{i}^{2}}-\mu}{T} \right ),
\end{equation}
with $\mu$ being the chemical potentials for the conserved quantities
and $T$ being the temperature. Unfortunately, most of the time this
ideal case is not realized during the evolution in the present classical
transport simulation. For the present investigation however, the
following criteria are more relevant, because they allow to include the
deviations from equilibrium.

Using the momentum-space anisotropy to characterize the local kinetic
equilibrium, one observes from Fig.\ \ref{Paniso} that kinetic
equilibrium may only be reached after 10 fm/$c$ (see also section
\ref{ssec:STE}). The question arises, whether the larger anisotropies of
the momentum distribution before 10 fm/$c$ affect the extracted energy
densities that are of relevance for our studies. We overcome this
problem by employing an anisotropic energy-momentum tensor for the
extraction of the energy density, as detailed below. The approach to
local chemical equilibrium is more difficult to quantify. Approximate
chemical equilibrium may also only be reached towards the end of the
reaction. Here we take the chemical off-equilibrium situation into
account by extracting a pion chemical potential and employing
corresponding fugacity factors.

The findings of the present work are in line with previous detailed
studies comparing the particle yields and spectra of the UrQMD transport
approach at different times with those of the statistical model
\cite{BraunMunzinger:1995bp,Cleymans:1996cd, Andronic:2005yp} which
showed that it takes roughly 10\,fm/$c$ until local kinetic and chemical
equilibrium is approximately reached for nucleus-nucleus collisions at
SPS energies \cite{Bravina:1998pi,Bravina:1998it,Bravina:1999dh}.

In summary, to account for the non-equilibrium effects in the present
study, which obviously dominate large parts of the evolution, we will
use the following scheme: One (i) considers the pressure (respectively
momentum) anisotropies in each cell and applies an approach developed
for anisotropic hydrodynamics to extract the effective energy density
which is used for all further considerations, (ii) introduces an EoS
assuming thermal and chemical equilibrium and (iii) finally extracts a
pion chemical potential $\mu_{\pi}$ which is the non-equilibrium effect
with the largest impact on the thermal dilepton rates. These aspects
will be considered in detail in the following. However, note that step
(i) only accounts for deviations from the purely kinetic condition of
isotropic momentum distributions and (iii) accounts for the chemical
off-equilibrium of the pions only. Besides, when applying the EoS,
kinetic and chemical equilibrium in the end remain an assumption here as
in any macroscopic approach.

\subsubsection{Kinetic anisotropies}
When considering the kinetic properties of the system, one finds that
the underlying transport approach in parts depicts large deviations from 
pressure isotropy. This is especially important for the early stages of the
reaction due to the large initial longitudinal momenta carried by the
nucleons of the two nuclei traversing each other. In consequence, the
longitudinal pressure is much higher than the transverse pressure. To
handle this kinetic off-equilibrium situation, a description developed
for anisotropic hydrodynamics is employed \cite{Florkowski:2010cf} 
which allows for differing longitudinal and transverse pressures. 
In this case the energy-momentum tensor takes the form
\begin{equation}
T^{\mu \nu} = \left( \varepsilon  + P_{\perp}\right) u^{\mu}u^{\nu} -
P_{\perp} \, g^{\mu\nu} - (P_{\perp} - P_{\parallel}) v^{\mu}v^{\nu}. 
\label{Taniso}
\end{equation}
Here $\varepsilon$ is the energy density, $P_{\perp}$ and
$P_{\parallel}$ are the pressures perpendicular and in direction of the
beam, respectively; $u^{\mu}$ is the fluid four-velocity and $v^{\mu}$
the four-vector of the beam direction. To define realistic values for
energy density $\varepsilon$ and pressure $P$ in the energy-momentum
tensor we introduce an anisotropy parameter
\begin{equation} \label{anpar}
x = (P_{\parallel}/P_{\perp})^{3/4}
\end{equation} 
and apply a generalized equation of state \cite{Florkowski:2012pf}
according to the following relations
\begin{alignat}{2}
\label{eps-aniso}
\varepsilon_{\text{real}} &= \varepsilon / r(x), \\
P_{\text{real}} &= P_{\perp} / [r(x)+3xr'(x)],  \\
P_{\text{real}} &= P_{\parallel} / [r(x)-6xr'(x)].
\end{alignat}
The relaxation function $r(x)$, with its derivative $r'(x)$,
characterizes the properties of a system which exhibits a Boltzmann-like
pressure anisotropy. It is given by
\begin{equation}
r(x) =\begin{cases} \label{relaxfunc}
        \frac{x^{-1/3}}{2}\left(1+\frac{x \artanh
        \sqrt{1-x}}{\sqrt{1-x}}\right) \text{for } x \leq 1  \\ 
        \frac{x^{-1/3}}{2}\left(1+\frac{x \arctan
        \sqrt{x-1}}{\sqrt{x-1}}\right) \text{for } x \geq 1
    \end{cases} .
\end{equation} 
With this procedure one can translate the anisotropic momentum
distribution into a local-equilibrium description that gives a realistic
value of $\varepsilon$ for our further calculations. This effective
model to account for the anisotropic pressure of the cell properties
allows to treat the early stage of the reaction in the same way as at
later times, when a local kinetic equilibration of the system has set
in. However, large differences between the ``regular'' energy density
$\varepsilon = T^{00}$ and the effective density
$\varepsilon_{\mathrm{real}}$ only show up for the first few
$\text{fm}/c$ of the evolution of the nuclear reaction. After that time
we find the longitudinal and perpendicular pressures being of at least
the same order of magnitude and
$\varepsilon_{\mathrm{real}} \approx \varepsilon$, i.e., no significant
deviations from the assumption of isotropic momentum distributions (for details see
Section \ref{ssec:STE} and \cite{Bravina:1999dh}). For all further studies
we assume that $\varepsilon_{\mathrm{real}}$ represents the energy density 
of the cell.

\subsubsection{Equation of state}
Having determined the rest-frame energy and baryon density, an equation
of state (EoS) is needed as additional input to calculate the
temperature $T$ and baryo-chemical potential $\mu_{\mathrm{B}}$ for each
cell. As the actual EoS for QCD matter is still not completely
determined, this is an uncertainty within the calculation. For the
present study we use a hadron-gas EoS with vacuum masses and without
mean-field potentials (HG-EoS) \cite{Zschiesche:2002zr, Petersen:2008dd}
following from hadronic chiral model calculations
\cite{Papazoglou:1998vr, Zschiesche:2006rf}. The included hadrons agree
with the degrees of freedom in UrQMD. However, this approach does not
account for a phase transition to a deconfined phase as it is neither
implemented in UrQMD nor in the hadron gas EoS. For heavy-ion collisions
at SPS energies, where we find initial temperatures significantly above
the expected critical temperature $T_{c}$, it will be also important to
consider dilepton emission from cells with such high temperatures during
the evolution of the medium. Therefore, we supplement the hadron-gas EoS
with a partonic equation of state \cite{He:2011zx} which is obtained
from a fit to lattice-QCD data of the form
\begin{equation}
\varepsilon/T^{4}=\frac{c\cdot(1+e^{-a/b})}{1+e^{(T_{c}/T-a)/b}}\cdot e^{\lambda T_{c}/T},
\end{equation}
with fit parameters $a=0.9979$, $b=0.1163$, $c=16.04157$ and
$\lambda=0.1773$ and critical temperature $T_{c}=170$\,MeV. To ensure a
smooth transition between the two EoS without any jumps in temperature
(i.e., to avoid discontinuities in the evolution), the values of $T$ from
the lattice EoS are matched with the HG-EoS in the temperature range
from 150-170\,MeV and exclusively used above $T_{c}$.

\begin{figure}[t]
\includegraphics[width=1.02\columnwidth]{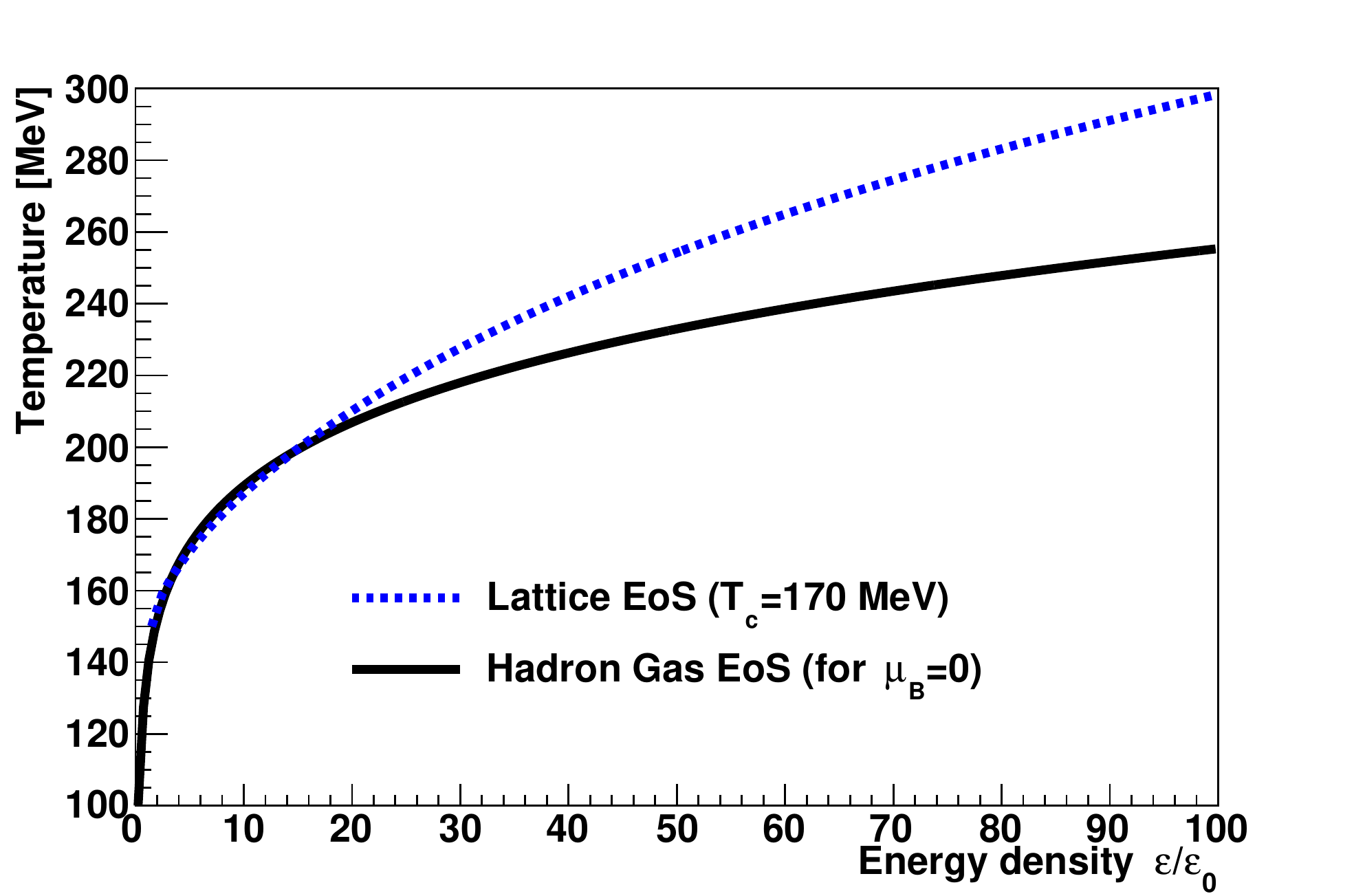}
\caption{\label{EoS} (Color online) Comparison of the two equations of
  state used in our model, the hadron gas EoS \cite{Zschiesche:2002zr}
  with UrQMD-like degrees of freedom (black line) and the Lattice EoS
  \cite{He:2011zx} to model the deconfined phase (blue line). We show
  the temperature dependence on the energy density (in units of normal
  nuclear-matter density, $\varepsilon_0$). While for the Lattice EoS
  $\mu_{\mathrm{B}}=0$ intrinsically, we set the chemical potential to
  zero as well for the hadron gas EoS for reasons of comparison.}
\end{figure}

In Figure \ref{EoS} a comparison of the relation between temperature and
energy density is shown for the two equations of state used in the
model. The hadron-gas EoS is represented by the black line and the
lattice EoS by the blue dashed line. While for the lattice EoS
$\mu_{\mathrm{B}}=0$ is valid intrinsically, the baryon density and in
consequence the chemical potential are set to zero in this plot for the
hadron gas EoS for reasons of comparison. However, the rather moderate
baryon chemical potential found at top SPS energy does not have 
a large impact on the relation between energy density and 
temperature and gives rise to a deviation from the curve for
$\rho_{\mathrm{B}}=\mu_{\mathrm{B}}=0$ of at maximum a few MeV. As
observed from Fig.\,\ref{EoS}, both equations of state agree very well
up to an energy density of roughly $15\varepsilon_{0}$, which
corresponds to a temperature of 200\,MeV. This implies that both EoS are
dual in the region around the phase transition, guaranteeing a smooth
cross-over transition when changing from the QGP EoS to the hadron-gas
EoS across the phase transition.

The parametrization of the Rapp-Wambach spectral function
\cite{RappSF}, which will be used in this study for convenient and
reliable application, is constructed such that the presence of baryonic
matter enters via a dependence on an effective baryon density
\begin{equation}
  \rho_{\text{eff}}=\rho_{\mathrm N} + \rho_{\bar{\mathrm N}} + 0.5 
  \left(\rho_{ \mathrm B^{*}} + \rho_{\bar{\mathrm  B}^{*}} \right).
\end{equation}
It includes nucleons and excited baryons as well as their
anti-particles. The reason not to take $\mu_{\mathrm B}$ is that the
interaction between the $\rho$ and a baryon is the same as with an
anti-baryon, i.e., it is not the net-baryon number that affects the
electromagnetic current-current correlation function but the sum of
baryons and anti-baryons. In our approach, we calculate the value of
$\rho_{\text{eff}}$ not via the EoS but directly from the cell's
rest-frame. 

\subsubsection{Pion chemical potential}

In full chemical equilibrium all meson chemical potentials are zero
since the meson number is not a conserved quantity. When applying the
EoS as described above, the explicit assumption is that in each cell we
find a thermally and chemically equilibrated system.  However, it has
been shown that in transport models during the initial non-equilibrium 
stage -- which is dominated by high energy densities -- an over-dense 
pionic system is created and remains for significant time-scales due to the
long relaxation time of pions \cite{Kataja:1990tp,Bebie:1991ij}. This is
mainly caused by initial string fragmentation and resonances decaying
into more than two final particles (e.g. $\omega \rightarrow 3\pi$) for
which the back-reaction channel is not implemented. Also macroscopic
approaches find non-zero $\mu_{\pi}$ after the number of pions is fixed
at the chemical freeze-out but the system further cools down and expands
\cite{Kolb:2002ve}. The appearance of a finite $\mu_{\pi}$ has a large
effect on the creation of $\rho$ mesons and therefore on the thermal
dilepton emission (see section \ref{ssec:Emission}). Therefore, though
in general assuming chemical equilibrium, we exclude the pions from this
assumption and extract a pion chemical potential in each cell via a
Boltzmann approximation. The according relation for a relativistic gas
is given by \cite{Sollfrank:1991xm}
\begin{equation}
  \mu_{\pi}= T\cdot\ln \left( \frac{2\pi^{2}n_{\pi}}{g_{\pi}Tm^{2}
      \mathrm{K}_{2}\left(\frac{m}{T}\right)} \right),
\end{equation}
with the pion density $n_{\pi}$ in the cell, the pion degeneracy
$g_{\pi}=3$, and the Bessel function of the second kind,
$\mathrm{K}_{2}$. Other meson chemical potentials as, e.g., a kaon
chemical potential are not considered in the present study, as the
dependence of the $\rho$ spectral function with regard to $\mu_{K}$ is
negligible.
\subsection{\label{ssec:Emission} Dilepton emission rates }

By assuming that the cells in our (3+1)-dimensional space-time grid are
in local equilibrium (except for the finite $\mu_{\pi}$) we can 
calculate the thermal emission from these cells. The dilepton 
emission is related to the imaginary part of the electromagnetic 
current-current correlation function \cite{Feinberg:1976ua,McLerran:1984ay},
$\im \Pi^{(\mathrm{ret})}_{\mathrm{em}}$. The full expression for the
dilepton emission rate per four-volume and four-momentum from a heat
bath at temperature $T$ and chemical potential $\mu_{\mathrm{B}}$ takes
the form \cite{Rapp:2013nxa}
\begin{equation}
\begin{split}
  \frac{\dd N_{ll}}{\dd^4x\dd^4q} = & -\frac{\alpha_\mathrm{em}^2 L(M)}{\pi^3 M^2} \;
  f^{\text{B}}(q \cdot U;T) \; \\
& \times \im \Pi^{(\text{ret})}_\mathrm{em}(M, \vec{q};\mu_B,T),
\end{split}
\label{rate}
\end{equation}
where $f^{\text{B}}$ denotes the Bose-distribution function and $L$ the lepton
phase-space factor,
\begin{equation} 
\label{phsp}
L(M)=\sqrt{1 - \frac{4m_{\mu}^{2}}{M^{2}}}
\left(1 +  \frac{2m_{\mu}^{2}}{M^{2}} \right),
\end{equation}
which reaches 1 rapidly above the threshold, given by twice the lepton mass.

To calculate invariant-mass spectra from equation (\ref{rate}) we 
integrate over four-volume and three-momentum
\begin{equation}
  \frac{\dd N_{ll}}{\dd M} = \int \dd^4x  \frac{M\dd^{3} \vec{p}}{p_{0}}\frac{\dd N_{ll}}{\dd^4x\dd^4p}
\end{equation}
In our case the integration over the four-volume simply reduces to a
multiplication of the cell's four-volume.

\subsubsection{Thermal $\rho$ emission}

There exist several approaches to calculate the in-medium spectral
functions, e.g., by using empirical scattering amplitudes
\cite{Eletsky:2001bb}. Here the $\rho$ spectral function from hadronic
many-body calculations by Rapp and Wambach \cite{Rapp:1999us} is used,
which has proven a good agreement with experimental results at
CERN-SPS and RHIC energies in previous studies
\cite{vanHees:2006ng,vanHees:2007th,Rapp:2013nxa}. In this approach, the
hadronic part of the electromagnetic current-current correlator is
saturated by light vector mesons according to the Vector Dominance Model
(VDM). The correlator is hereby related to the $\rho$ spectral function,
respectively the propagator, as
\begin{equation}
  \im \Pi_\mathrm{em}^{(\text{ret})}=\frac{(m^{(0)}_{\rho})^{4}}{g^{2}_{\rho}} \im D_{\rho}^{(\text{ret})}.
\end{equation}
To determine the propagator, several contributions to the self-energy
have to be considered, i.e., in this case the meson gas
($\Sigma_{\rho M}$) and nuclear matter effects ($\Sigma_{\rho B}$) as
well as the in-medium $\rho\pi\pi$ width ($\Sigma_{\rho\pi\pi}$). This
results in
\begin{equation}
D_{\rho}=\frac{1}{M^{2}-(m_{\rho}^{(0)})^{2}-
\Sigma_{\rho\pi\pi}-\Sigma_{\rho M}-\Sigma_{\rho B}}.
\end{equation}
For the present study we use a parametrization of the Rapp-Wambach
spectral function \cite{RappSF} that has been checked against the full
spectral function and proven to agree well, with a maximal deviation of
up to 15\% in the mass region around 0.4 GeV.

To arrive at the final yield $N_{\rho \rightarrow ll}$ we have to
generalize (\ref{rate}) for a chemical off-equilibrium state with finite
pion chemical potential. It is necessary to include an additional
(squared) fugacity factor, which is in Boltzmann approximation
\begin{equation}
z^{2}_{\pi}=\exp \left(\frac{2\mu_{\pi}}{T}\right).
\end{equation}
The reason for this is that the above expression for the dilepton
emission rate (\ref{rate}) is independent of the hadronic initial and
final states as only chemical potentials of conserved charges are
considered (for which $Q_{i} = Q_{f}$). However, since the pion number
$N_{\pi}$ is not a conserved quantity this assumption is no longer
correct for $\mu_{\pi}\neq 0$ which means that generally $N_{\pi,i} -
N_{\pi,f} \neq 0$.
\subsubsection{Multi pion emission}
While the $\rho$ meson is expected to dominate the dilepton emission in
the low invariant mass range up to $M=1 \, \GeV$, a continuum starts to
develop above, and multi-pion interactions will contribute at higher
masses.

As detailed in \cite{Dey:1990ba}, the vector correlator at finite
temperature takes the form
\begin{equation}
\Pi^{V}_{\mu\nu}(p,T)=(1-\varepsilon)\Pi_{\mu\nu}^{V}(p,0)+\varepsilon
\Pi_{\mu\nu}^{A}(p,0) 
\end{equation} 
with the mixing parameter $\varepsilon=T^{2}/(6F_{\pi}^{2})$ ($F_{\pi}$:
pion-decay constant). In this paper, we follow the same approach as
presented in \cite{vanHees:2006ng,vanHees:2007th}, using
\begin{equation}
\begin{split}
  \Pi_{V}(p) = &(1-\varepsilon)z_{\pi}^{4}\Pi^{\text{vac}}_{\text{V},4\pi} +
  \frac{\varepsilon}{2}z^{3}_{\pi} \Pi^{\text{vac}}_{\text{A},3\pi} \\
  & + \frac{\varepsilon}{2}(z^{4}_{\pi}+z^{5}_{\pi})\Pi^{\text{vac}}_{\text{A},5\pi}
\end{split}
\end{equation} 
according to vector/axial-vector correlators from tau-decay data
provided by ALEPH \cite{Barate:1998uf}. The pion chemical potentials are
implemented via the fugacity factor $z_{\pi}$. The two-pion piece, as
well as the three-pion piece corresponding to the $a_{1}$ decay,
$a_{1} \rightarrow \pi + \rho$, have been excluded as they are already
included via the $\rho$ spectral function.

\subsubsection{Quark-gluon plasma emission}
\begin{figure}[t]
\includegraphics[width=1.02\columnwidth]{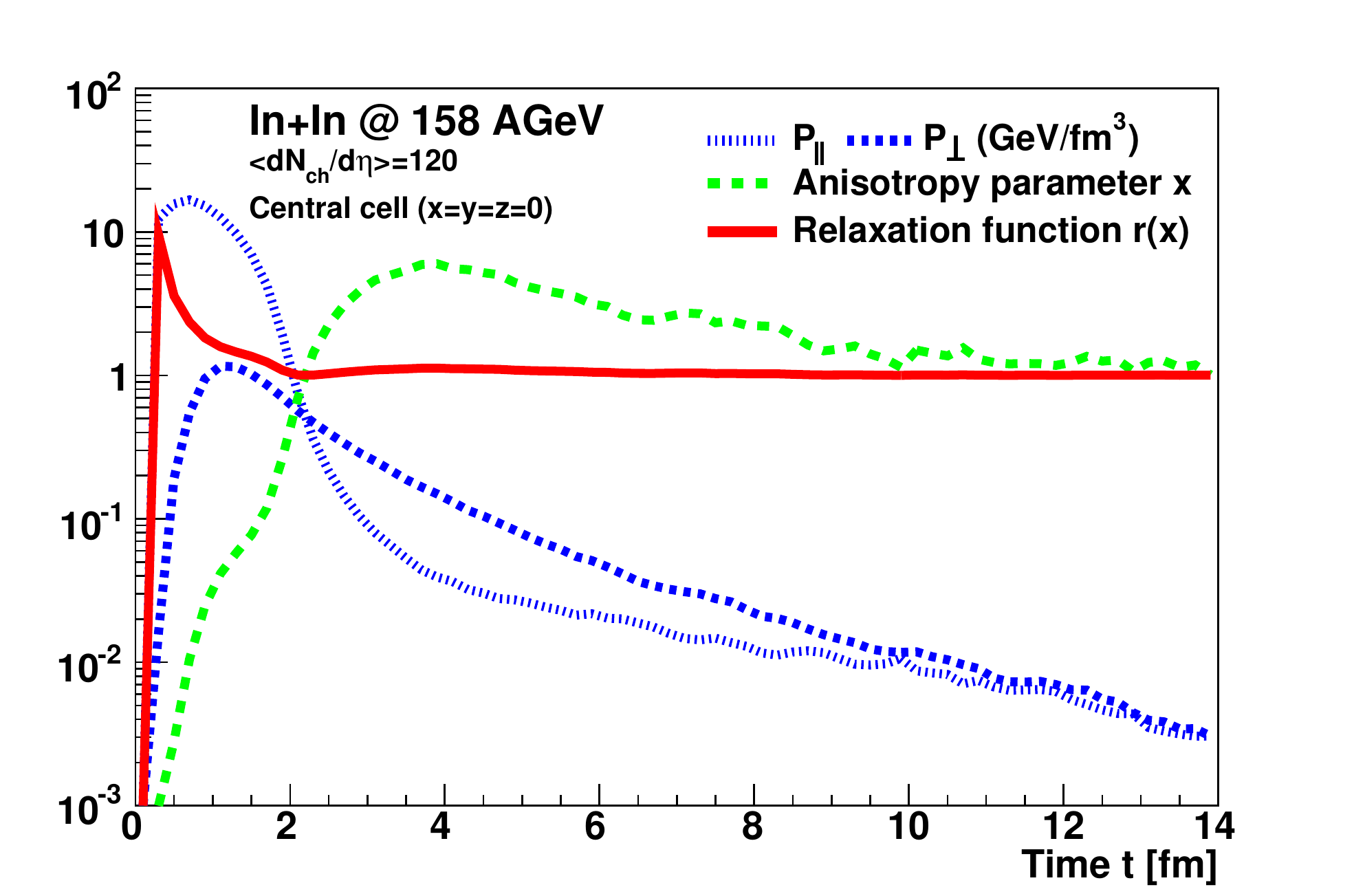}
\caption{\label{Paniso} (Color online) Longitudinal and transverse
  pressure for the central cell, as well as the anisotropy parameter
  $x = (P_{\parallel}/P_{\perp})^{3/4}$ and the relaxation function
  $r(x)$ as defined in equation (\ref{relaxfunc}).}
\end{figure}
\begin{figure*}[t]
\includegraphics[width=1.02\columnwidth]{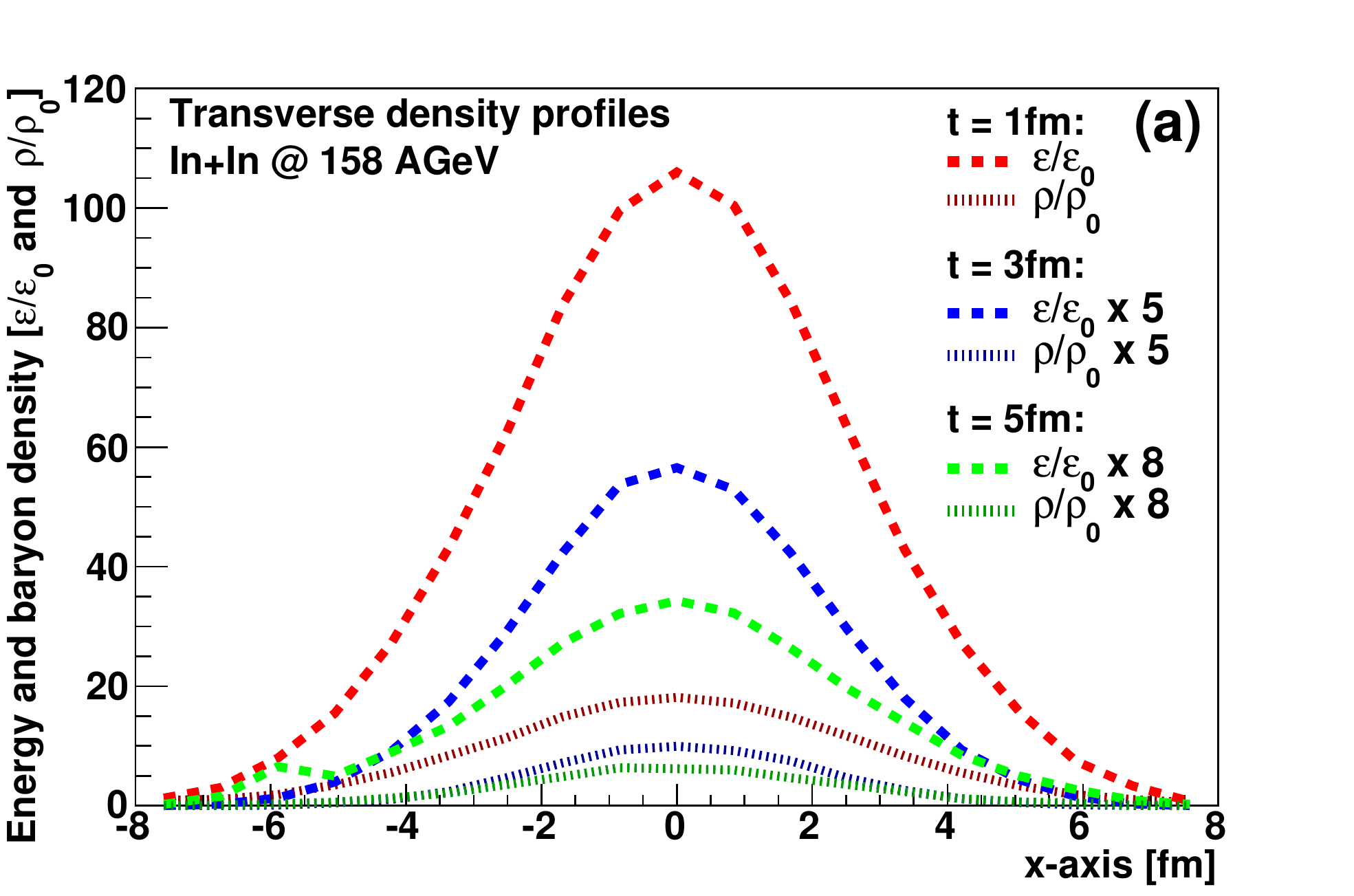}
\includegraphics[width=1.02\columnwidth]{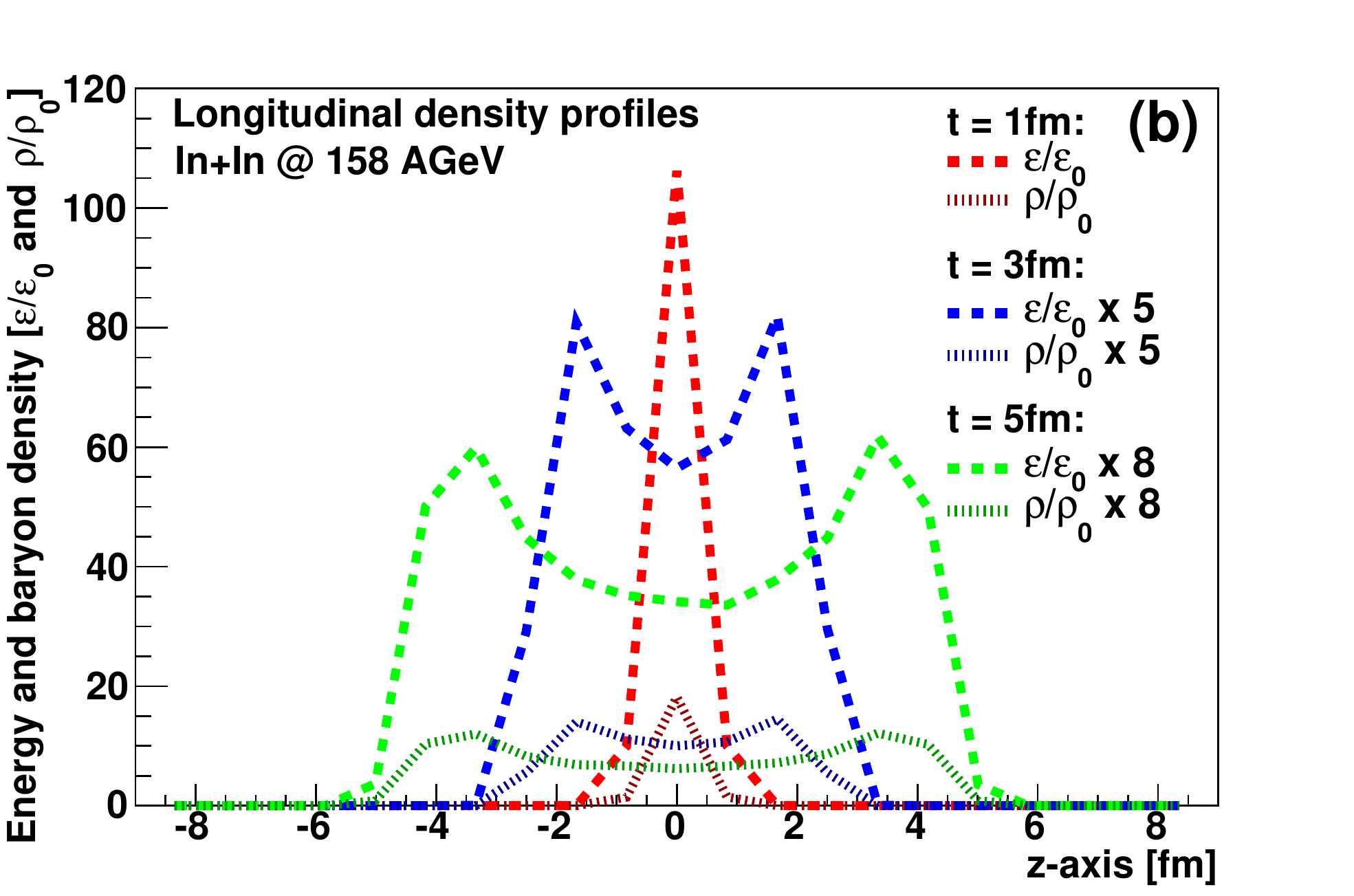}
\caption{\label{profile} (Color online) Transverse (a) and longitudinal
  (b) profiles of the energy density $\varepsilon$ and the baryon
  density $\rho_{ \mathrm{B}}$. The left plot shows the dependence of
  $\varepsilon$ and $\rho_{\mathrm{B}}$ on the position along the
  $x$-axis, with $y$ and $z$ coordinates fixed to 0. The right plot
  shows the dependence along the $z$-axis (i.e., along the beam axis)
  with $x=y=0$. The results are presented in units of the normal
  nuclear-matter densities, $\varepsilon_{0}$ and $\rho_{0}$. }
\end{figure*}
The dilepton emission from the quark-gluon plasma has been considered to
be one of the most promising probes for the formation of a deconfined
phase. In such a QGP phase, a quark can annihilate with an anti-quark
into a dilepton pair (via a virtual photon).

Here we use a corresponding spectral function extrapolated from lattice
QCD correlators for three-momentum $\vec{q}=0$ \cite{Ding:2010ga} and
with a light-like limit, consistent with the leading order $\alpha_{s}$
photon production rate \cite{Rapp:2013nxa}. The emission rate per
four-volume and four-momentum takes the form
\begin{equation} \frac{\dd N_{ll}}{\mathrm{d^{4}}x{\dd^{4}}p} = \frac{\alpha_{\mathrm{em}}^{2}}{6\pi^{3}} \sum\limits_q e_q^2  \frac{\rho_{\mathrm{V}}(p_{0},\vec{p},T)}{(p_{0}^{2}-\vec{p}^{2})(\ee^{p_{0}/T}-1)} ,
\end{equation} 
where $\rho_{V}$ denotes the vector spectral function.  The current
calculation assumes that the chemical potential is zero in the
deconfined phase, i.e., that the quark and anti-quark distributions are
equal.

For comparison also the pure perturbative quark-gluon plasma
contribution is calculated. The rate has been evaluated for lowest order
$q\bar{q}$ annihilation \cite{Cleymans:1986na} as
\begin{equation}
\begin{split} 
  \frac{\dd N_{ll}}{\dd^4x\dd^4p} = & \frac{\alpha_{em}^2}{4\pi^4}
  \frac{T}{p} f^{\text{B}}(p_0;T) \sum\limits_q e_q^2 \\
  & \times \ln \frac{\left(x_-+y\right) \left(
      x_++\exp[-\mu_q/T]\right)} {\left(x_++y\right) \left(
      x_-+\exp[-\mu_q/T]\right)}
\label{qqrate2} 
\end{split}
\end{equation}
with the expressions $x_\pm=\exp[-(p_0\pm p)/2T]$ and
$y=\exp[-(p_0+\mu_q)/T]$. Again, the quark chemical potential $\mu_{q}$
is zero for our considerations.

\subsubsection{Non-thermal $\rho$ emission}
In addition to the thermal contribution, we also have to handle those
cells where (i) the temperature is lower than 50 MeV (late stage), i.e.,
where it is not reasonable to assume a thermal emission and (ii) with no
baryon but only meson content (in these cells the density is usually
also quite low) which inhibits a determination of a local rest frame
according to the Eckart description and in consequence one can not
determine $T$ and $\mu_{ \mathrm{B}}$ in this way. In these cases we
directly take the $\rho^{0}$ mesons from the UrQMD calculations. Within
the transport approach they are mainly produced either via decay of
heavy resonances (e.g. $N^{*}_{1520}\rightarrow \rho N$) or the reaction
$\pi\pi \rightarrow \rho$. Production via strings is possible as
well. For these $\rho^{0}$ mesons we apply a shining procedure that is
conventionally used to calculate dilepton emission from a transport
approach \cite{Schmidt:2008hm}.

The mass-dependent width for the direct decay of a $\rho^{0}$ meson to a
dilepton pair is expressed according to \cite{Li:1996mi}
\begin{equation} \label{direct}
\Gamma_{V \rightarrow ll}(M) = \frac{\Gamma_{V \rightarrow ll}(m_{\rho})}{m_{\rho}} \frac{m_{\rho}^{4}}{M^{3}} \cdot L(M),
\end{equation}
with the partial decay width at the $\rho$-pole mass
$\Gamma_{\rho \rightarrow ll}(m_{\rho})$; $L(M)$ denotes the lepton
phase-space factor (\ref{phsp}).

The according dilepton yield is then obtained by summing over all
$\rho^{0}$ mesons from the low temperature cells,
\begin{equation}\label{shining}
  \frac{dN_{ll}}{dM} =\frac{\Delta N_{ll}}{\Delta M}= \sum _{i=1}^{N _{\Delta M}}\sum _{j=1}^{N _{\rho}}\frac{\Delta t}{\gamma_{\rho} } \frac{\Gamma_{\rho \rightarrow ll}(M)}{\Delta M}.
\end{equation}
Here $\Gamma _{\rho \rightarrow ll}(M)$ is the electromagnetic decay
width of the considered resonance defined in (\ref{direct}) and
$\Delta t$ is the length of a time step within our calculation. The factor
$\gamma_{\rho}^{-1}$ is introduced to account for the fact that the
$\rho$ meson lives longer in the center of mass system of the UrQMD
calculation than in its rest frame in which the shining is applied
(relativistic time dilation).

Note that we use the shining procedure only for cells for which we do
\emph{not} calculate the thermal emission, in consequence we avoid any
form of double-counting.

\section{\label{sec:Results} Results }
\begin{figure*}[t]
\includegraphics[width=1.0\columnwidth]{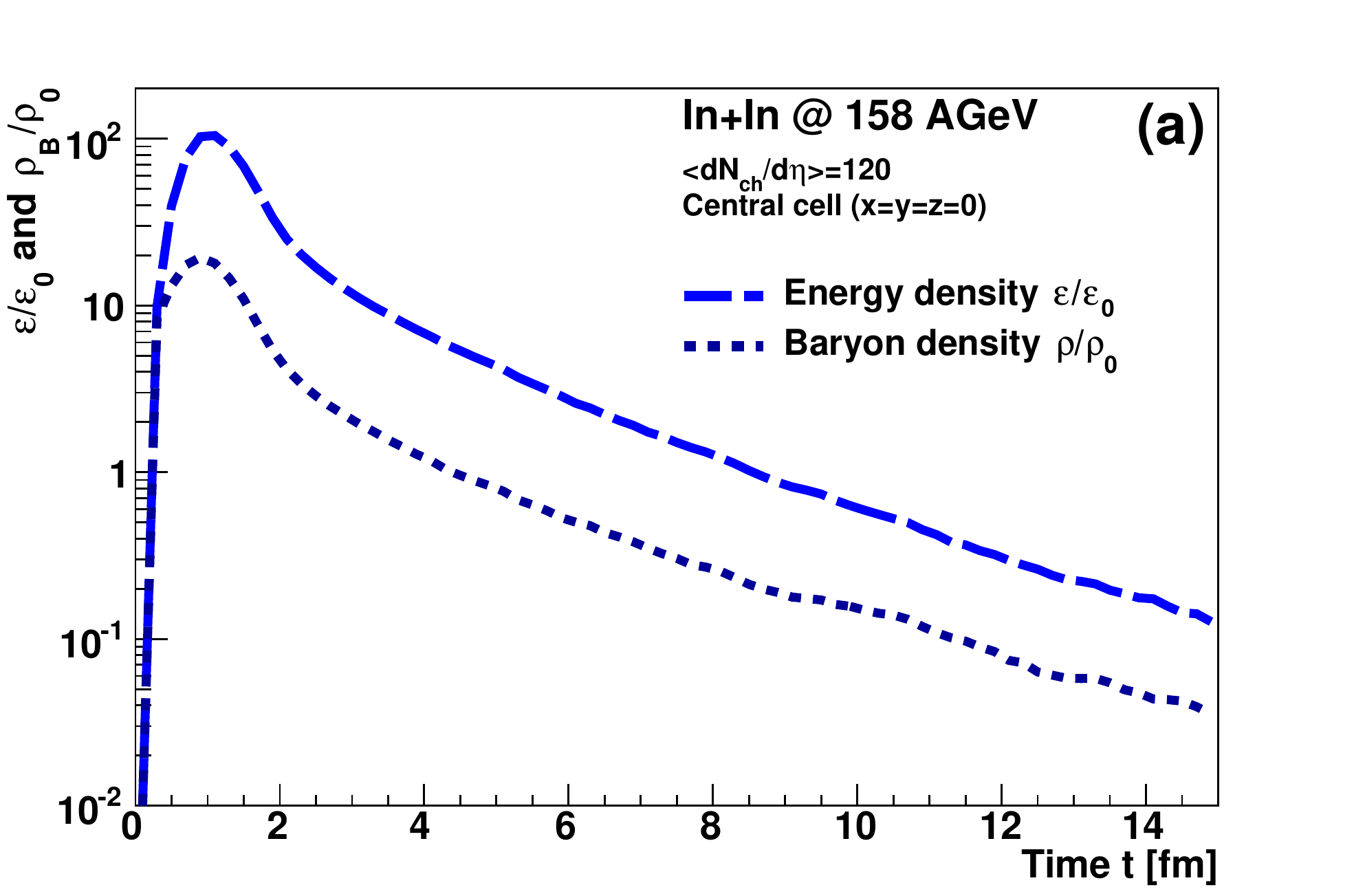}
\includegraphics[width=1.0\columnwidth]{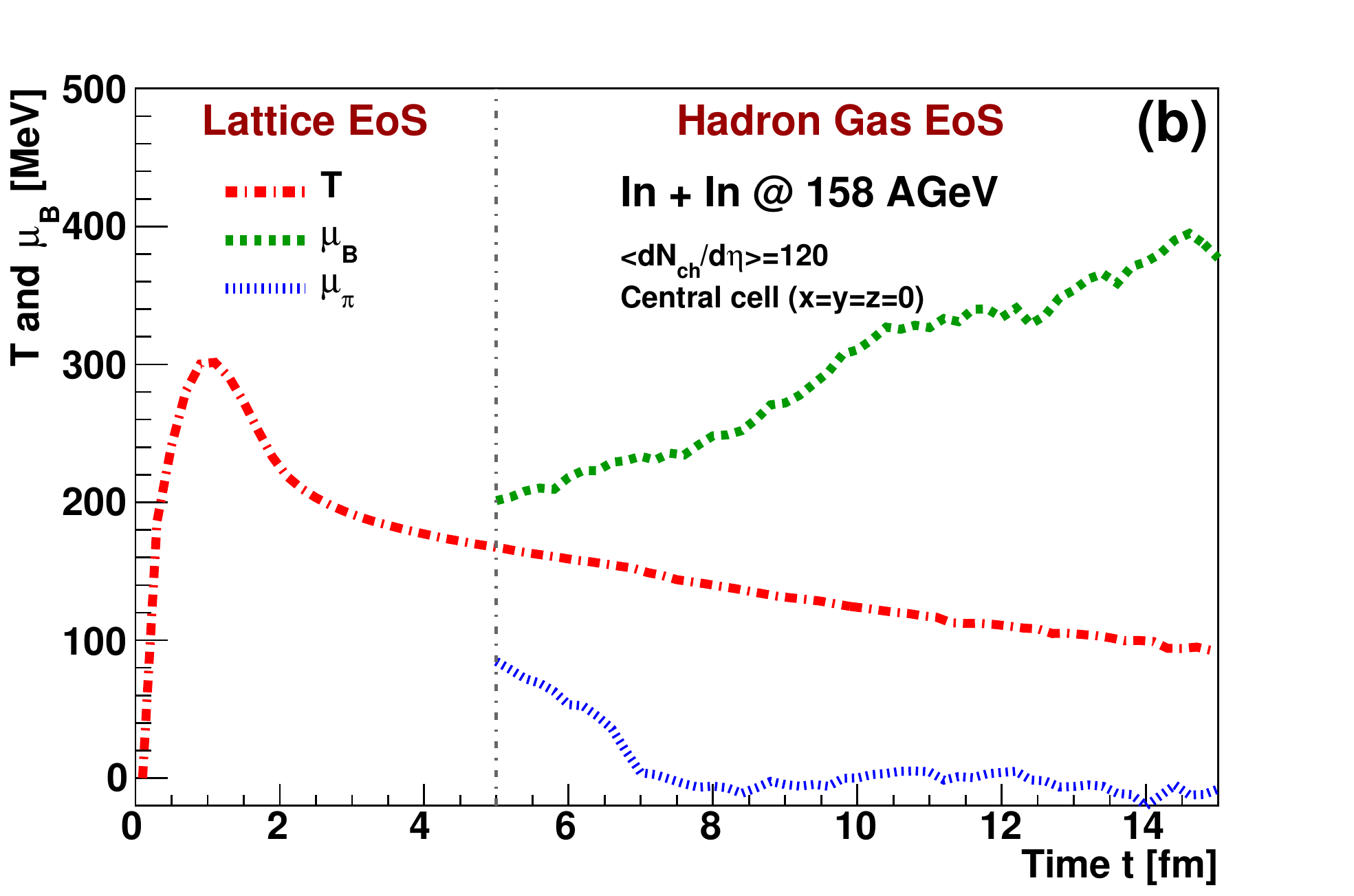}
\caption{\label{timeev} (Color online) Left panel (a): Time evolution of
  the baryon density $\rho_{\mathrm{B}}$ (short dashed) and energy
  density $\varepsilon$ (long dashed) for the cell at the center of the
  coarse-graining grid ($x=y=z=0$). The results are given in units of
  the ground-state densities $\varepsilon_{0}$ and $\rho_{0}$. Right
  panel (b): Time evolution of the temperature $T$ (red dash-dotted),
  baryon chemical potential $\mu_{\mathrm{B}}$ (green short dashed) and
  the pion chemical potential $\mu_{\pi}$ (blue dotted) in the central
  cell. The thin grey line indicates the transition from the Lattice EoS
  to the Hadron Gas EoS at the transition temperature of $T=170\,\MeV$.}
\end{figure*}
The following calculations were performed with the coarse-graining
approach as described above. To compare our results with the data
recorded by the NA60 Collaboration, an input of 1000 UrQMD events with a random
impact-parameter distribution restricted to $b<8.5\;\fm$ has been used,
which corresponds to a value of
$\erw{\dd N_{\mathrm{ch}}/\dd \eta} \approx 119$ in one unit of rapidity
around mid-rapidity in the center-of-mass frame. This is very close to
the value of $\erw{\dd N_{\mathrm{ch}}/\dd \eta}_{\mathrm{exp}} = 120$
which NA60 measured within their acceptance.

Note that several coarse-graining runs with different UrQMD events as
input had to be performed to obtain enough statistics, especially for
the non-thermal $\rho$ contribution.
\subsection{\label{ssec:STE} Space-time evolution }

As dileptons are radiated over the whole space-time evolution of the
nuclear reaction, it is important to model the dynamics as realistically
as possible. When studying the electromagnetic radiation from the hot
and dense phase of heavy-ion collisions, the thermodynamic properties of
the fireball are the main input for the calculations.

We first investigate the anisotropic situation at the beginning of the
collision. Figure \ref{Paniso} shows -- for the central cell at the
origin of our grid -- the time evolution of the longitudinal and
transverse pressure, $P_{\parallel}$ and $P_{\perp}$, together with the
anisotropy parameter, $x$, and the relaxation function, $r(x)$, as
defined in Eqs. (\ref{anpar}) and (\ref{relaxfunc}). As one expects, the
first time steps are characterized by high values of $P_{\parallel}$
while $P_{\perp}$ is rather negligible first but increases significantly
later. Both quantities initially differ by three orders of magnitude. In
the course of the evolution the values are approaching each other and
become equal around $t=2\,\fm/c$.  But they do not remain equal at later
times but the perpendicular pressure then exceeds the longitudinal one
by a factor 3-5 in the further development until they finally equalize
around $t=10\,\fm/c$. This finding agrees with previous studies of the
kinetic equilibration within the UrQMD model \cite{Bravina:1999dh}.
When considering the influence of the pressure anisotropy on the local
thermodynamic properties one has to bear in mind that according to
(\ref{eps-aniso}) the realistic energy density in the cell is determined
by the relaxation function $r(x)$. Its value drops rapidly from an
initial value of 10 (i.e. the realistic isotropized energy density is
here only 10\% of the actual energy density in the cell) to 1 at
$t=2$\,$\fm/c$ and remains on that level. In consequence, the anisotropy
does not significantly influence the thermodynamic properties in the
cell after 2\,$\fm/c$.

The transverse and the longitudinal profiles of the energy density
$\varepsilon$ and the baryon density $\rho_{B}$ are presented in Figure
\ref{profile}. The left plot (a) shows the distribution of $\rho$ and
$\varepsilon$ in dependence on the position along the $x$-axis whereas
$y=z=0$ and the right plot (b) shows the dependence on the $z$-position
(i.e. along the beam axis) with $x=y=0$. The results are plotted for
three different time steps, at 1, 3 and 5\,$\fm/c$ after the beginning
of the collision. (Note that the results for $t=3 \,\fm/c$ and
$5\,\fm/c$ are scaled up for better comparability.) The shape of the
transverse profile looks almost identical at all times with a clear peak
at $x=0$, which is falling off smoothly on both sides at the more
peripheral and less dense regions where consequently also less energy is
deposited. However, both energy and baryon density decrease clearly in
the course of time due to the expansion of the system. This expansion
can directly be seen in the longitudinal profile on the right side. On
the one hand, we see one clear peak structure at $t=1 \,\fm/c$ when the
two nuclei still mostly overlap. However, at $t=3 \, \fm/c$ and
$5 \, \fm/c$ we observe a double-peak structure with maxima that sheer
off from each other. A region of high $\varepsilon$ and
$\rho_{\mathrm{B}}$ is created in between when the two nuclei have
traversed each other. Stopping effects have produced this hot and dense
zone, whereas some remnants of the nuclei still fly apart with high
velocity.
\begin{figure*}[t]
\includegraphics[width=1.02\columnwidth]{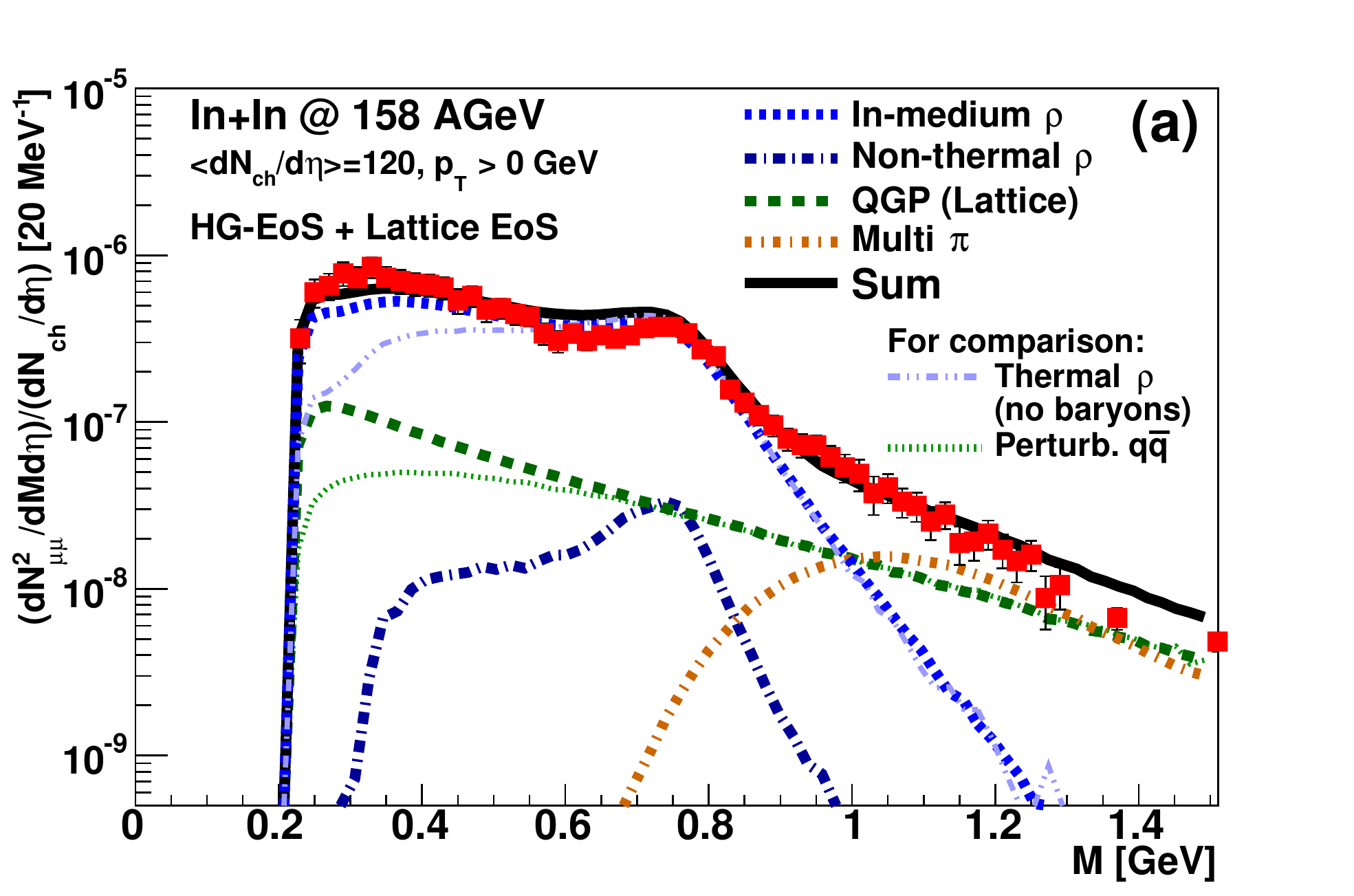}
\includegraphics[width=1.02\columnwidth]{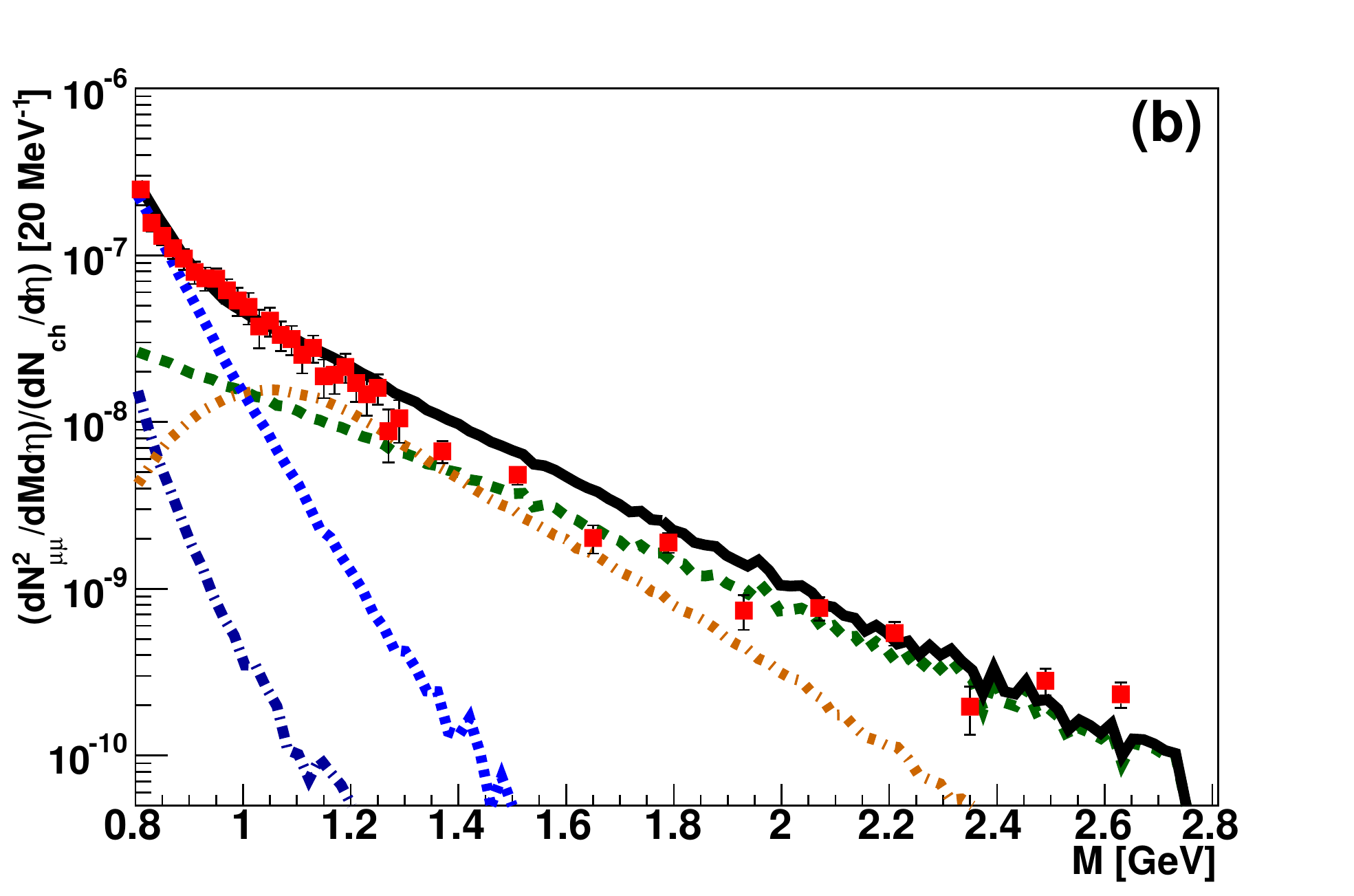}
\caption{\label{invmass} (Color online) Invariant mass spectra of the
  dimuon excess yield in In+In collisions at a beam energy of
  $158 \,A \GeV$, for the low-mass region up 1.5\,GeV (a) and the
  intermediate-mass regime up to $2.8\,\GeV$ (b). We show the
  contributions of the in-medium $\rho$ emission according to the
  Rapp-Wambach spectral function \cite{RappSF} (blue short dashed), the
  contribution from the Quark-Gluon Plasma, i.e.,
  $q\bar{q}$-annihilation, according to lattice rates
  \cite{Ding:2010ga,Rapp:2013nxa} (green dashed) and the emission from
  multi-pion reactions, taking vector-axial-vector mixing into account
  \cite{vanHees:2007th} (orange dash-dotted). Additionally a non-thermal
  transport contribution for the $\rho$ is included in the yield (dark
  blue dash-dotted). Only left plot: For comparison the thermal $\rho$
  without any baryonic effects, i.e. for $\rho_{\text{eff}}=0$, is shown
  (violet dash-double-dotted) together with the yield from pure
  perturbative $q\bar{q}$-annihilation rates (green dotted). The results
  are compared to the experimental data from the NA60 Collaboration
  \cite{Arnaldi:2008er, Specht:2010xu, Arnaldi:2008fw}}
\end{figure*} 
\begin{figure*}[t]
\includegraphics[width=1.02\columnwidth]{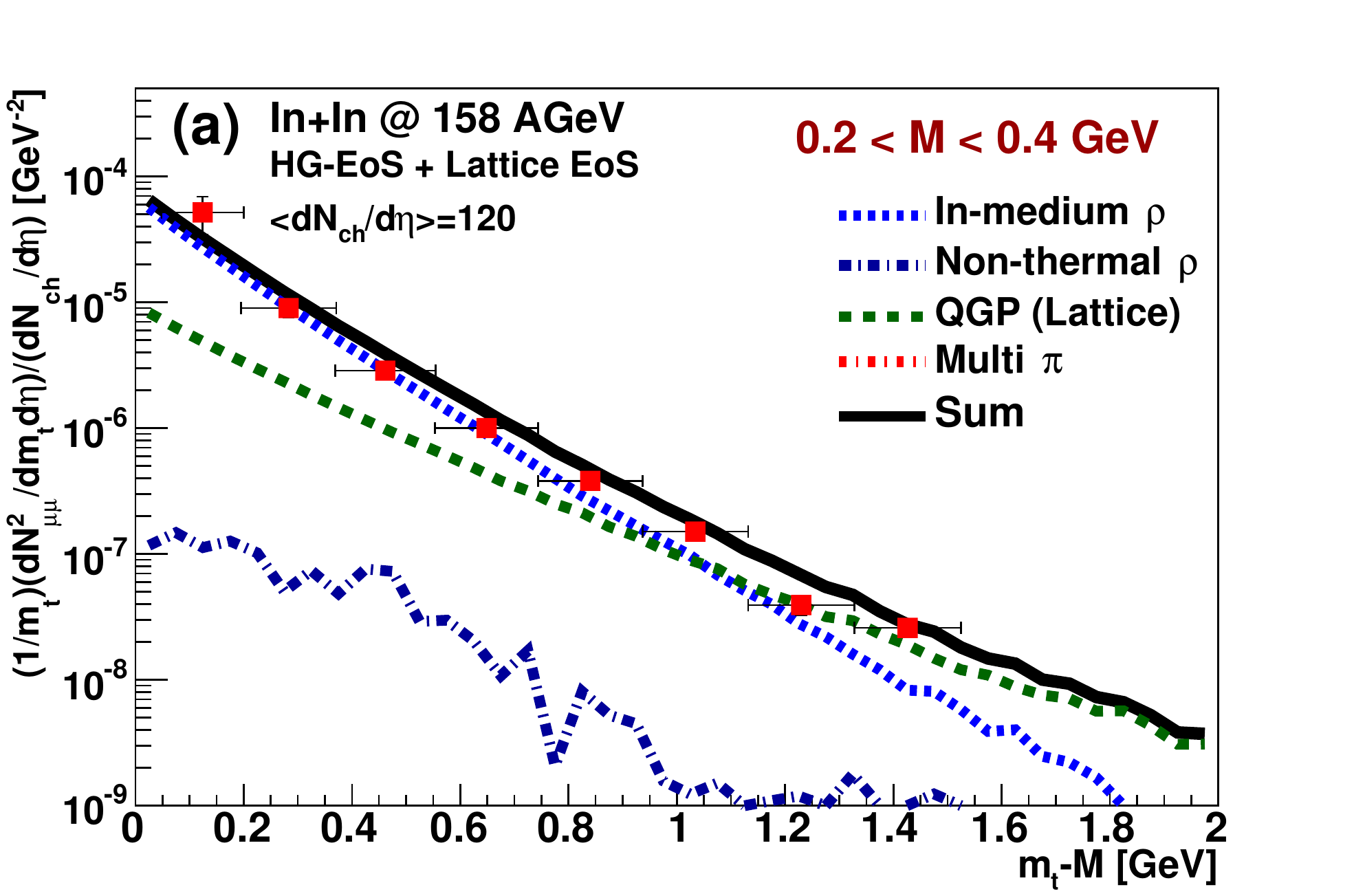}
\includegraphics[width=1.02\columnwidth]{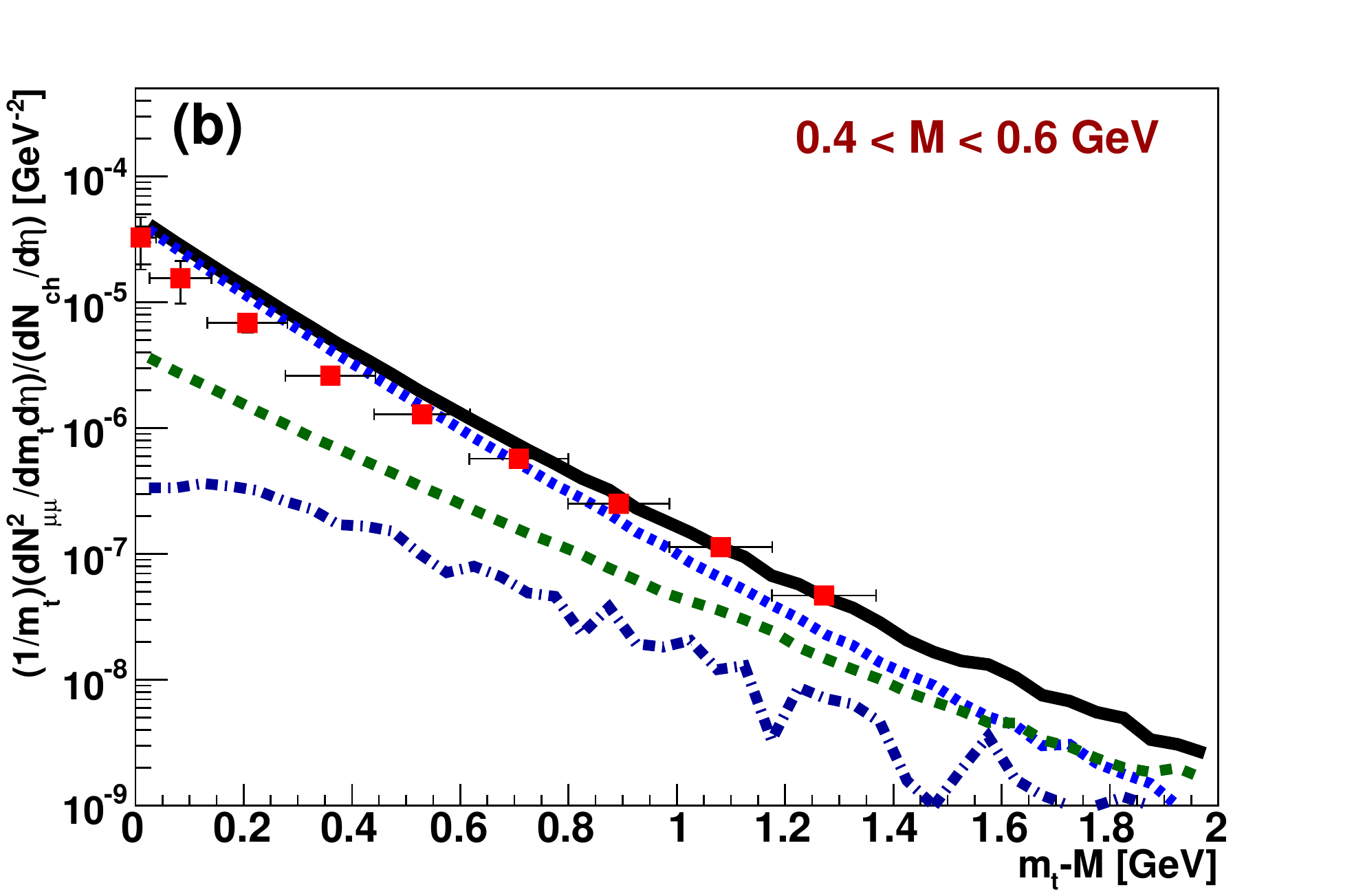}
\includegraphics[width=1.02\columnwidth]{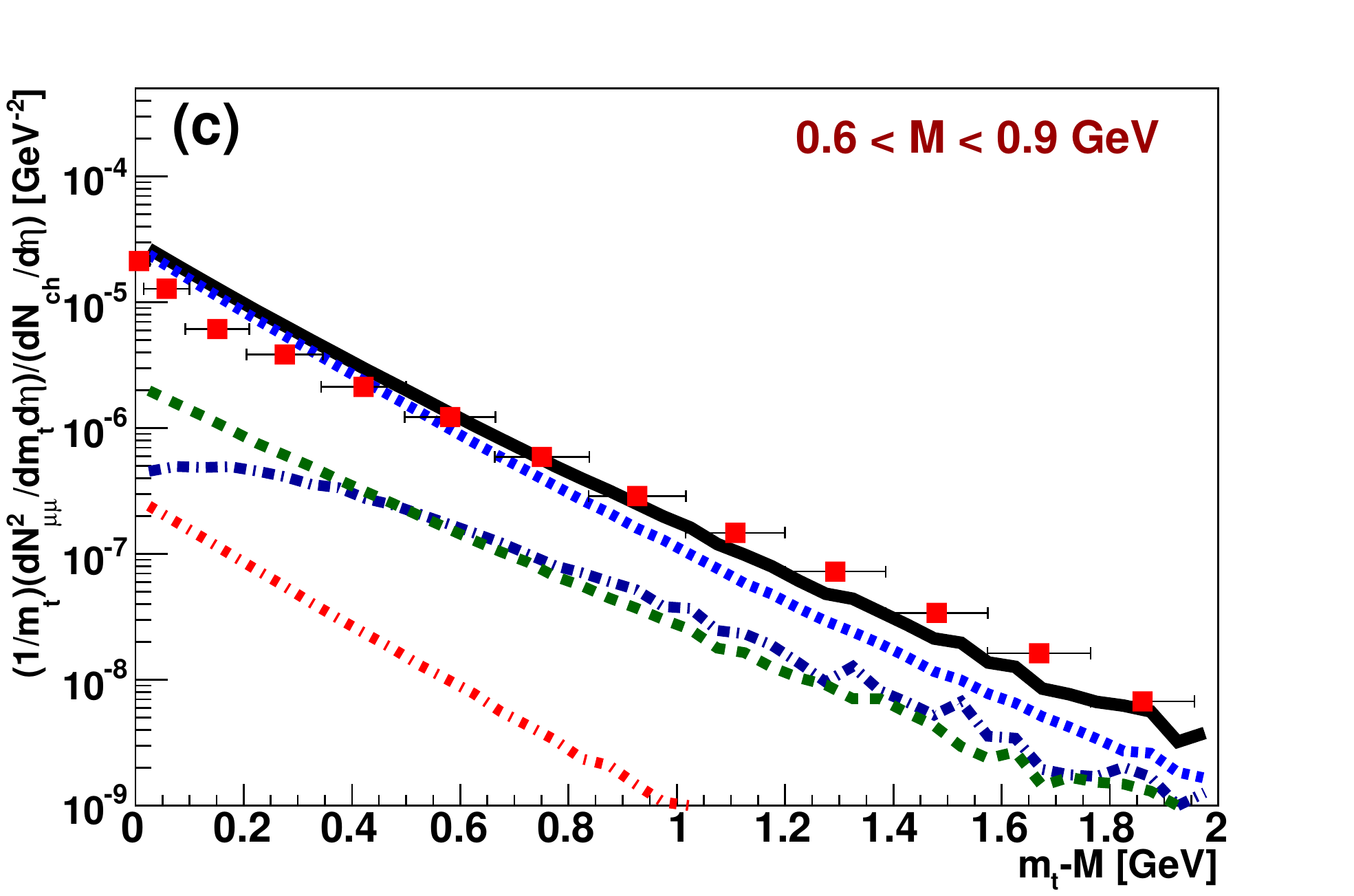}
\includegraphics[width=1.02\columnwidth]{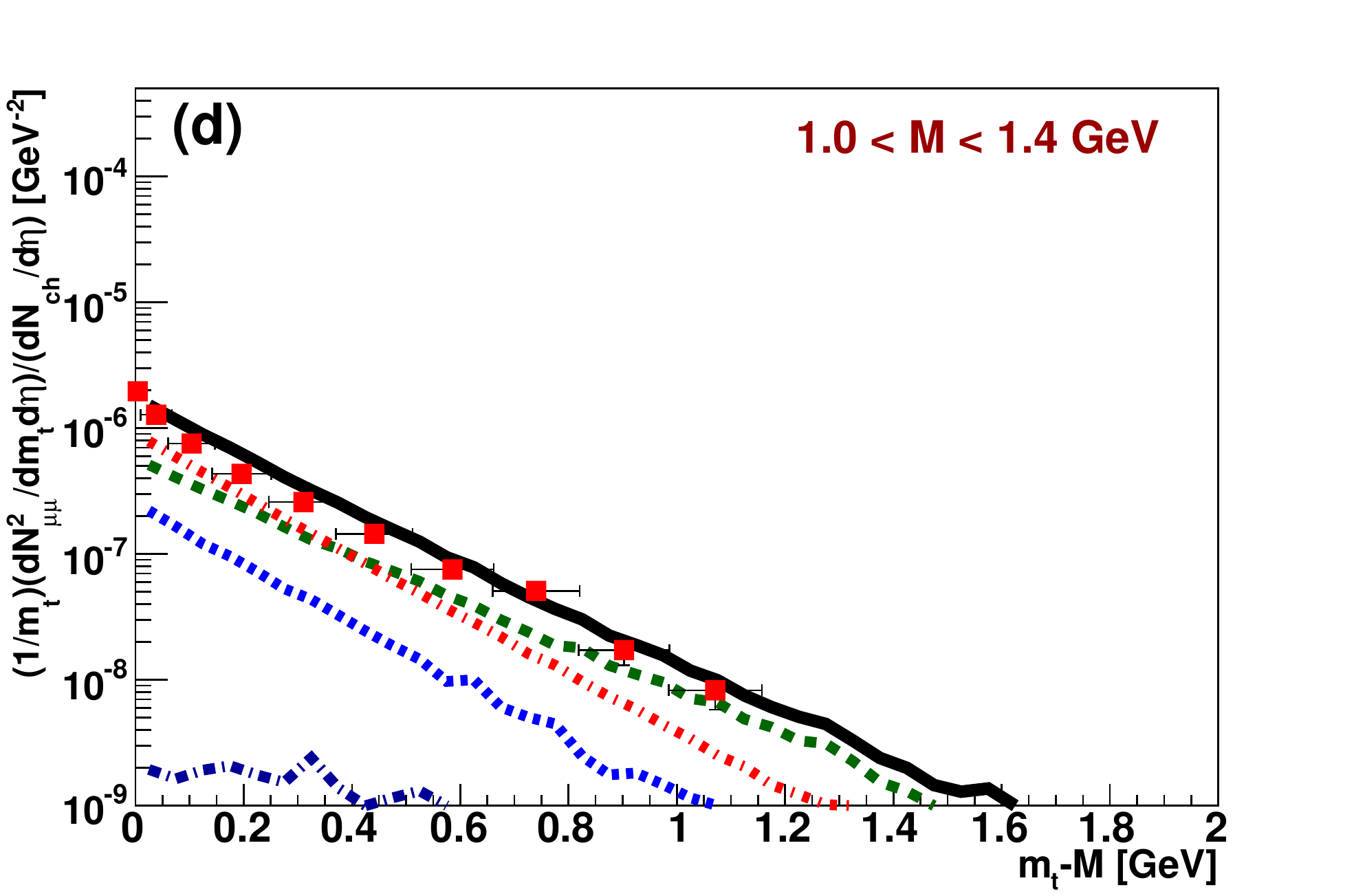}
\caption{\label{mtspectra} (Color online) Transverse-mass ($m_{t}-M$)
  spectra in four different mass bins of the dimuon-excess yield in
  In+In collisions at a beam energy of $158\, A \GeV$. We have the mass
  ranges $0.2 \, \GeV<M<0.4\,\GeV$ (a), $0.4\,\GeV<M<0.6\,\GeV$ (b),
  $0.6 \, \GeV<M<0.9\,\GeV$ (c) and the highest mass bin with
  $1.0 \, \GeV<M<1.4\,\GeV$ (d). The different contributions are the
  same as in Figure \ref{invmass}. The results are compared to the
  experimental data from the NA60 Collaboration
  \cite{Damjanovic:2007qm}.}
\end{figure*}
\begin{figure*}[t]
\includegraphics[width=1.02\columnwidth]{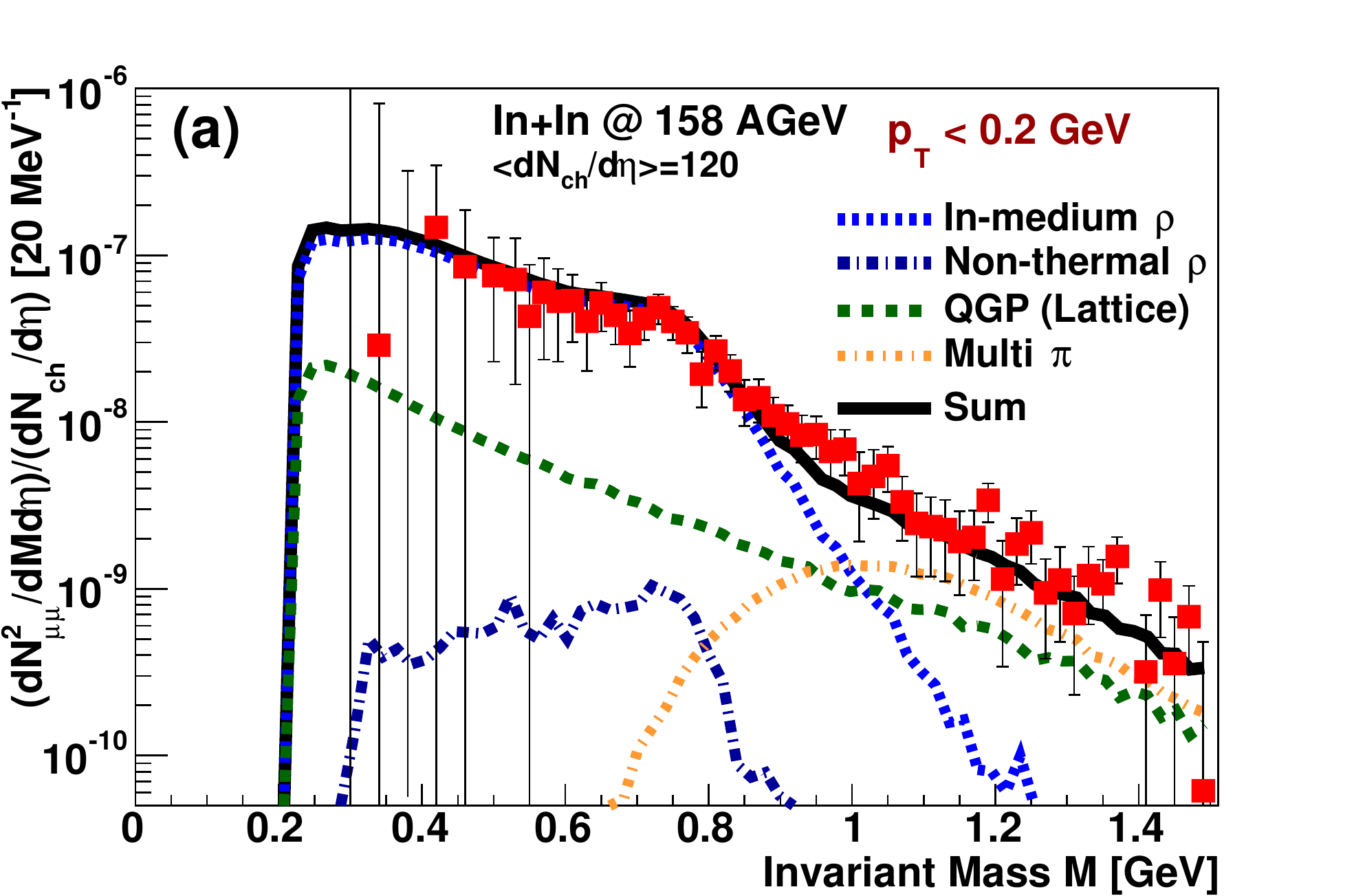}
\includegraphics[width=1.02\columnwidth]{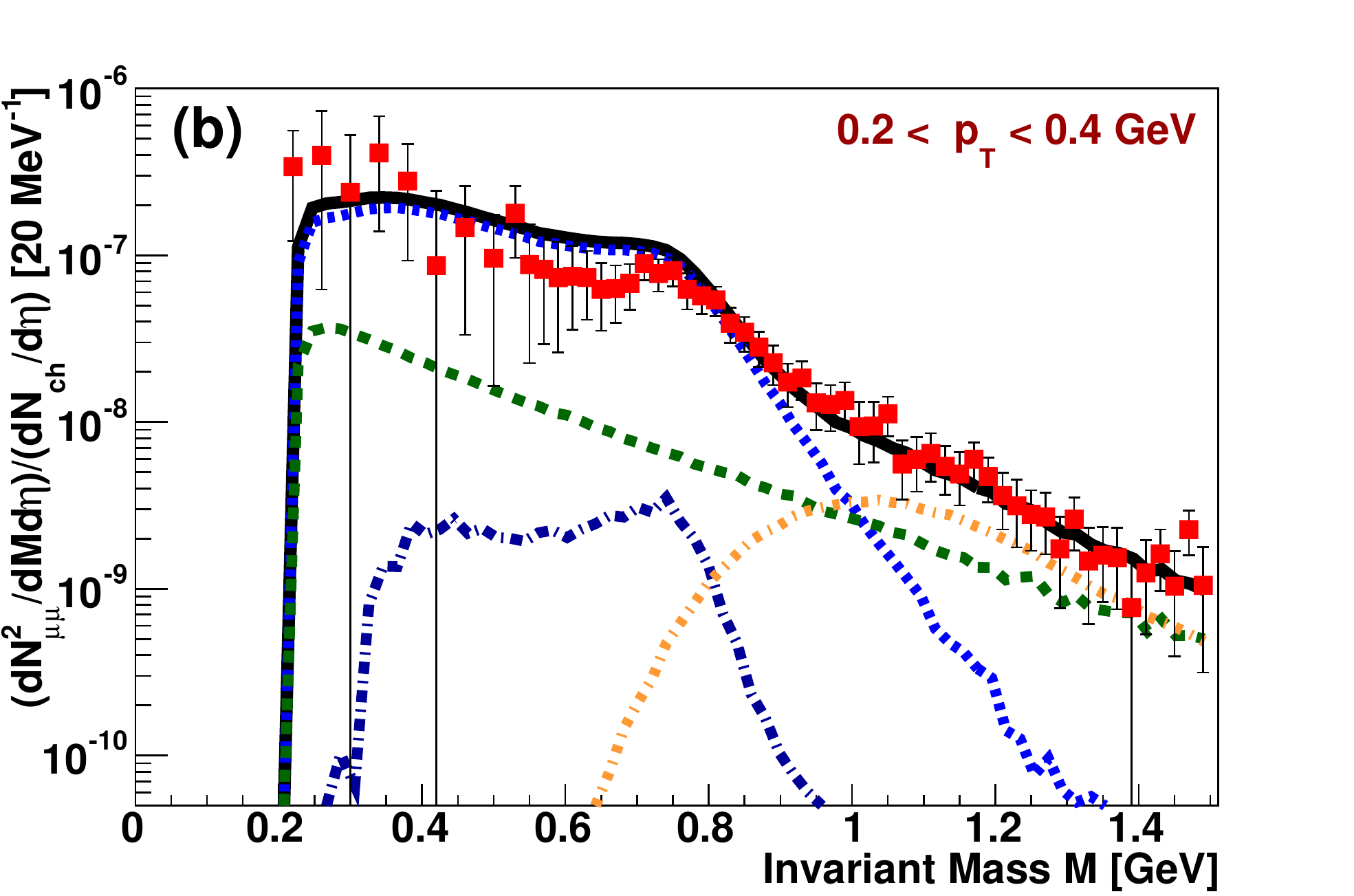}
\includegraphics[width=1.02\columnwidth]{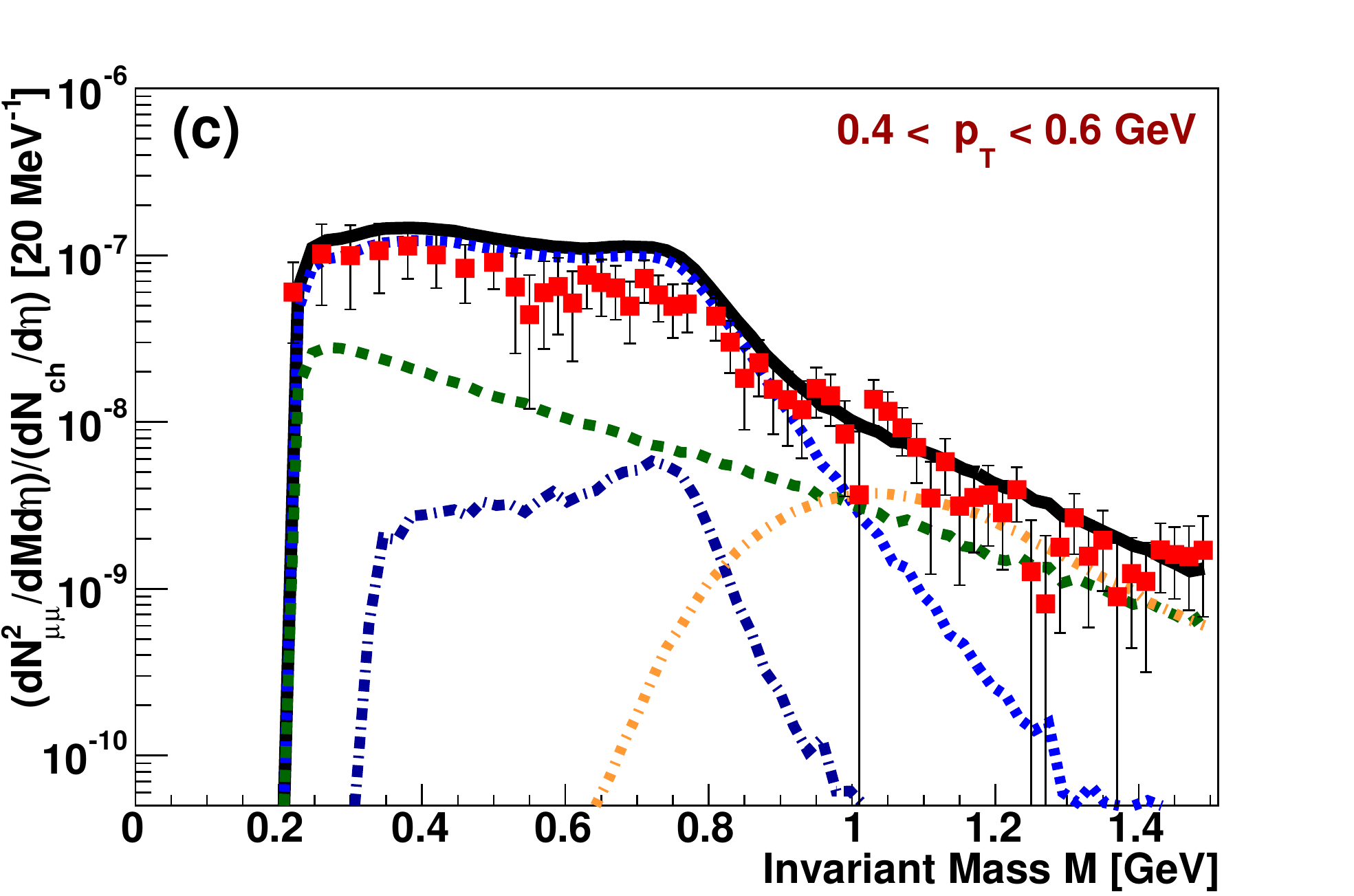}
\includegraphics[width=1.02\columnwidth]{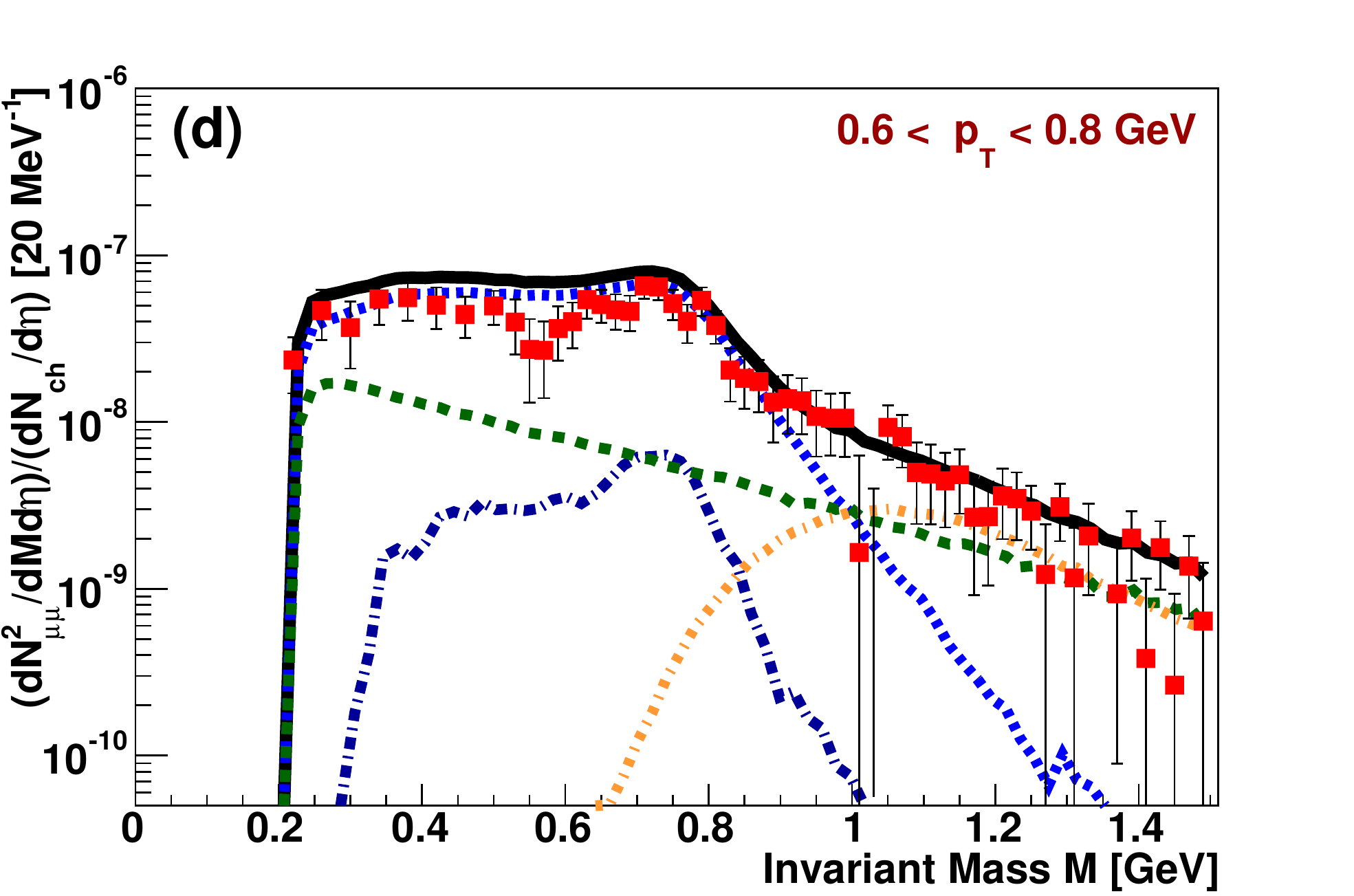}
\includegraphics[width=1.02\columnwidth]{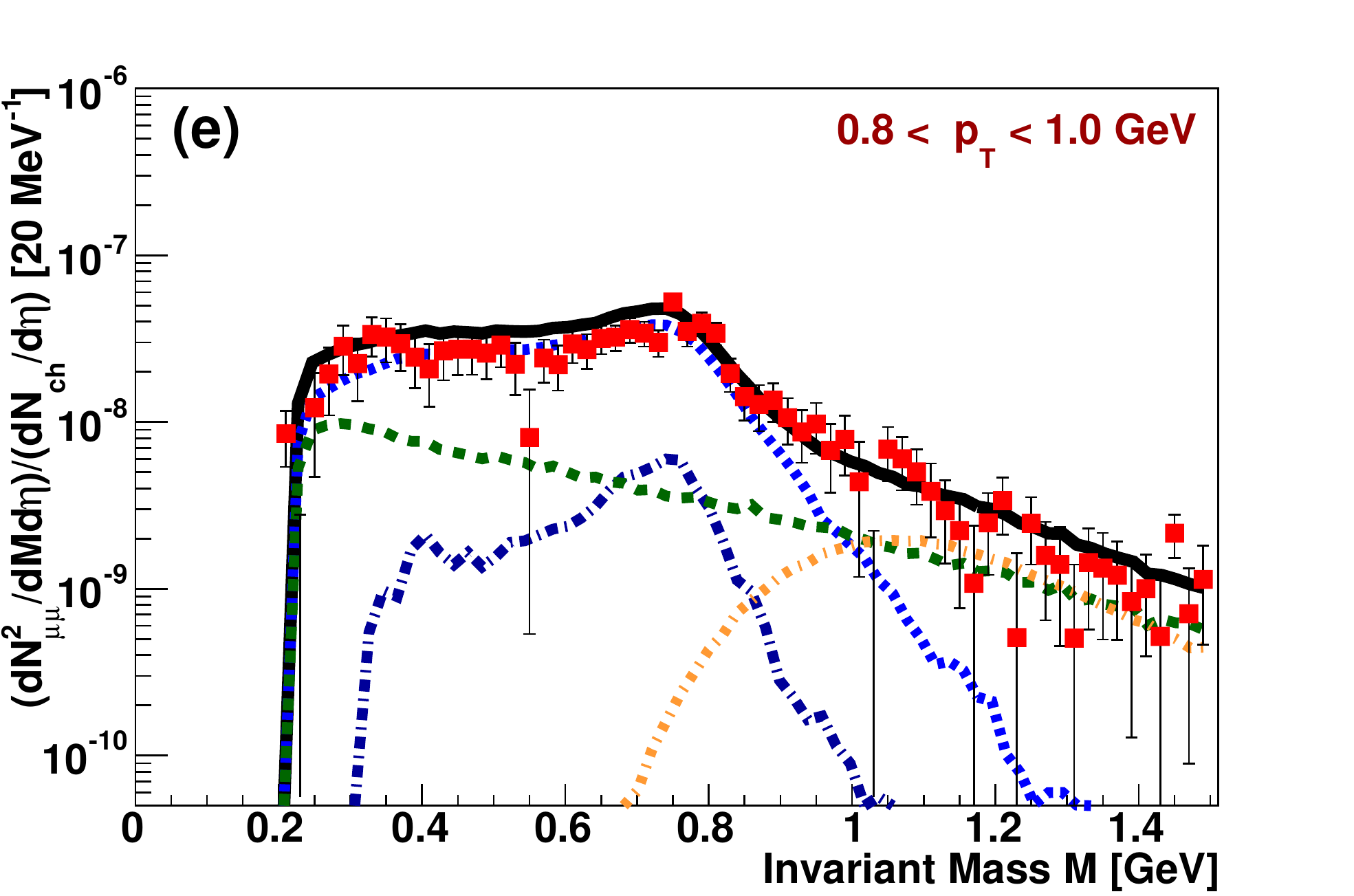}
\includegraphics[width=1.02\columnwidth]{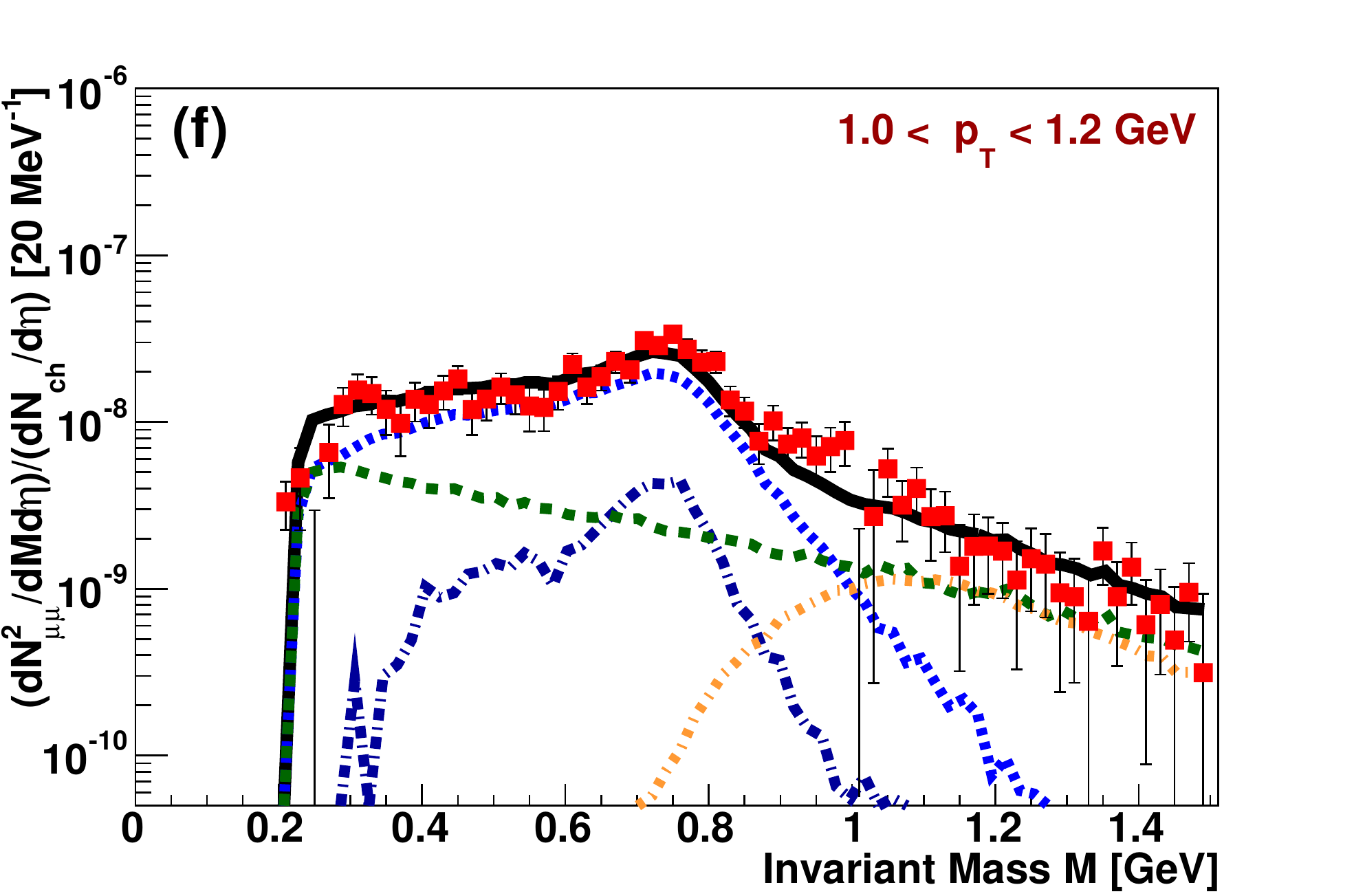}
\caption{\label{ptslices1} The invariant mass spectra of the dimuon
  excess yield in In+In collisions at a beam energy of $158\, A \GeV$,
  for the low-mass region as in the left plot of Figure \ref{invmass},
  but for 6 different $p_{t}$-bins ranging from $p_{t}=0$ to
  $1.2\,\GeV$. The different contributions are the same as in Figures
  \ref{invmass} and \ref{mtspectra}. The results are compared to the
  experimental data from the NA60 Collaboration \cite{Arnaldi:2008fw}.}
\end{figure*}
\begin{figure*}[t]
\includegraphics[width=1.02\columnwidth]{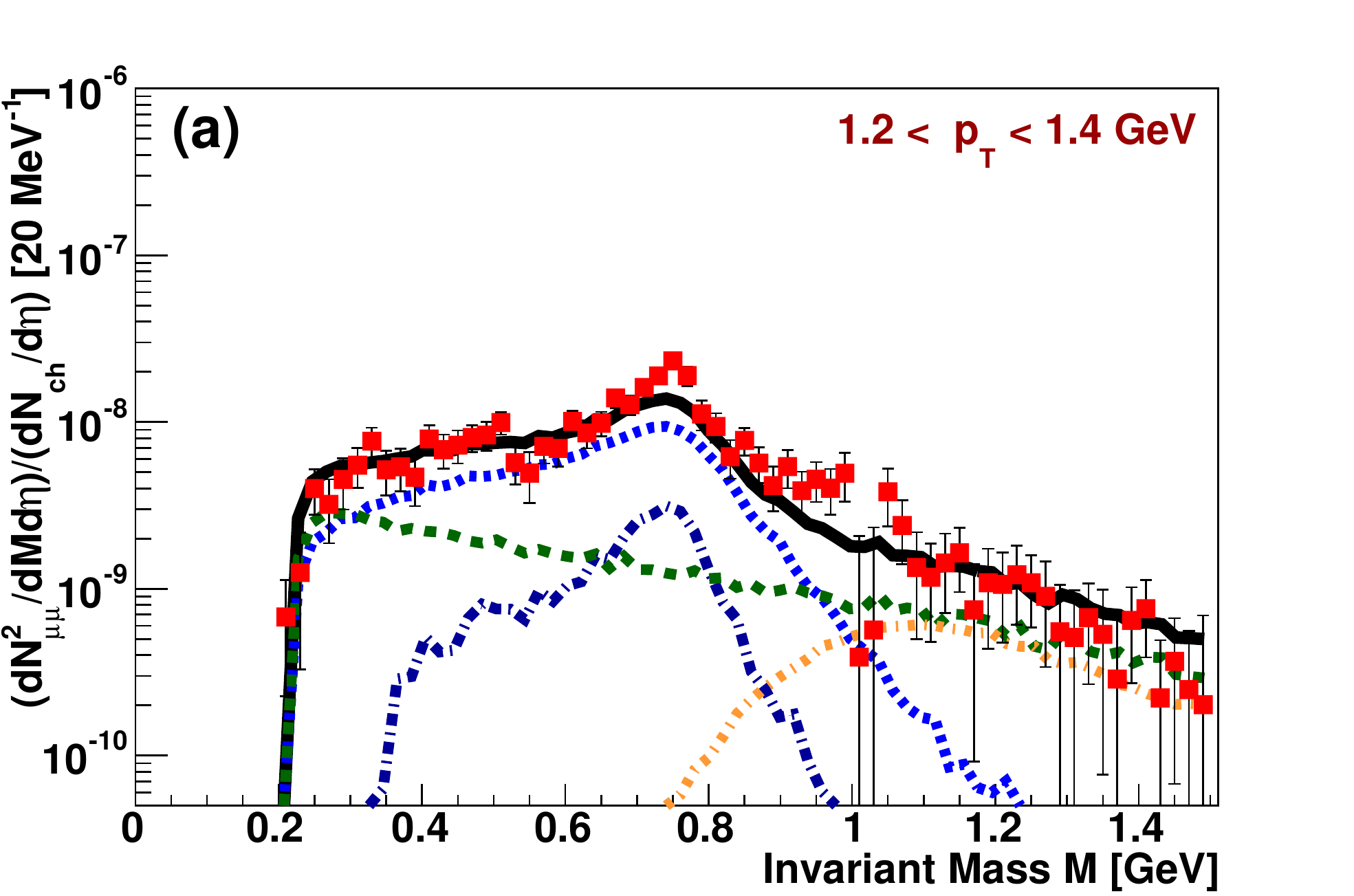}
\includegraphics[width=1.02\columnwidth]{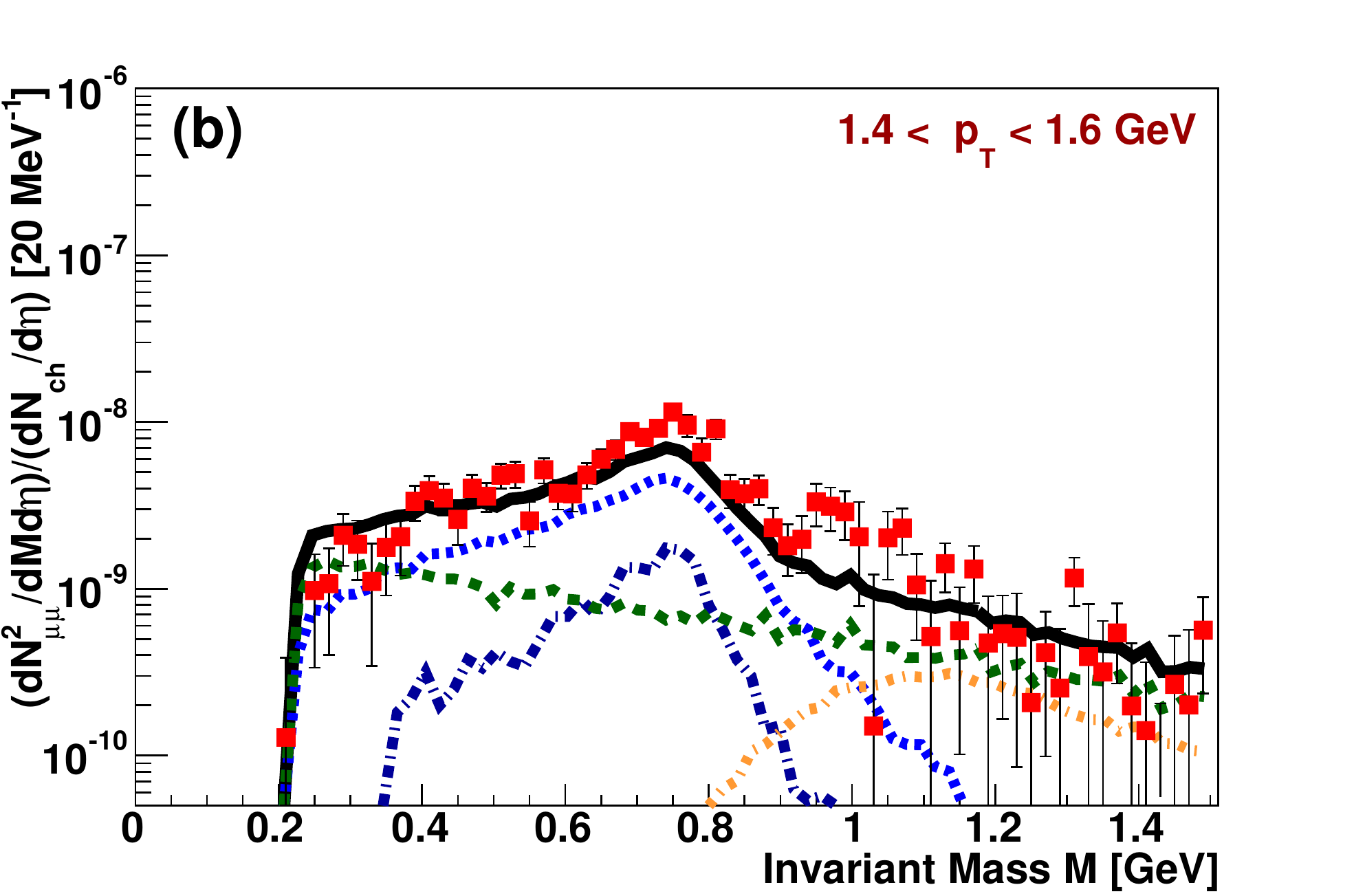}
\includegraphics[width=1.02\columnwidth]{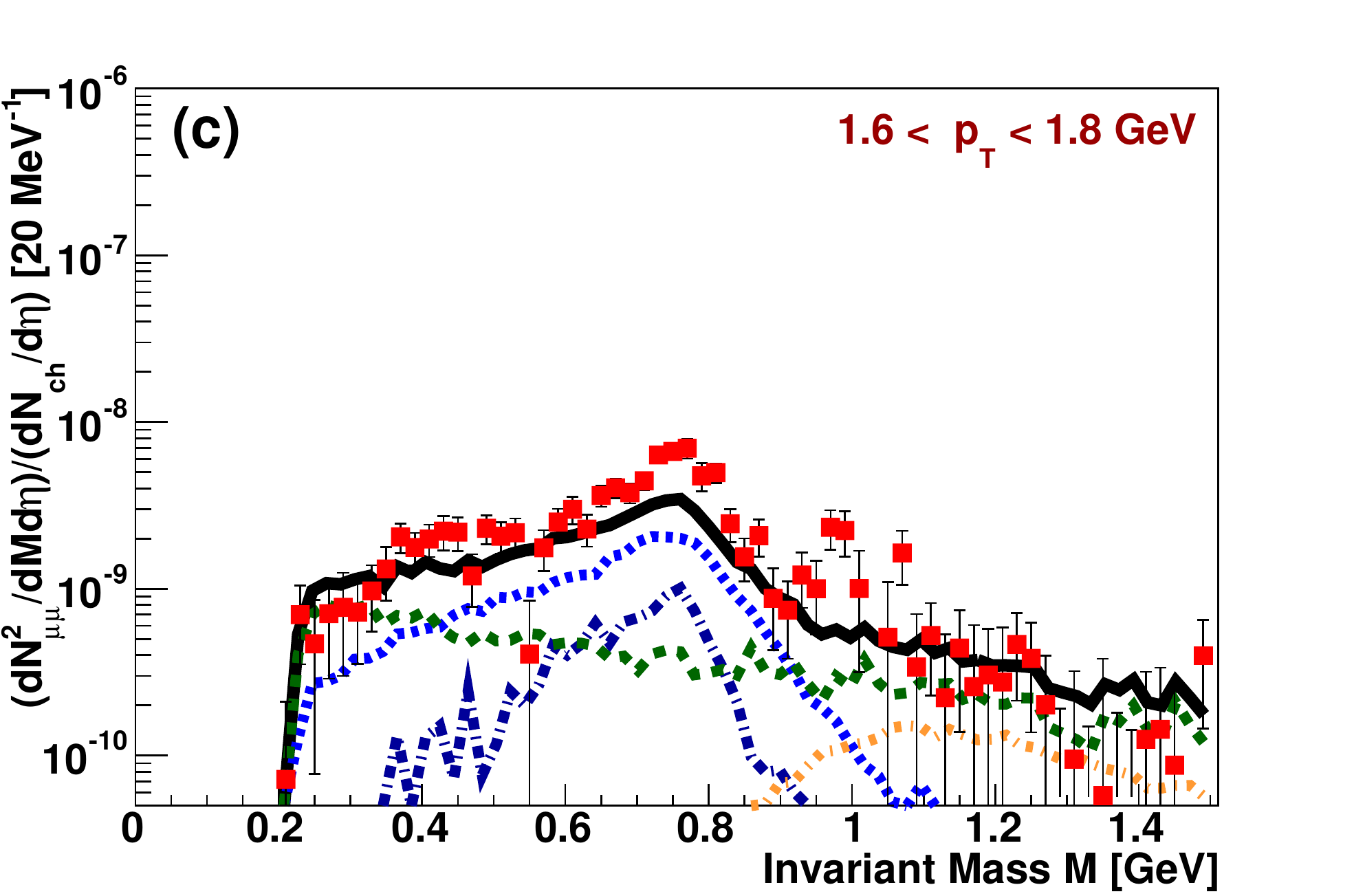}
\includegraphics[width=1.02\columnwidth]{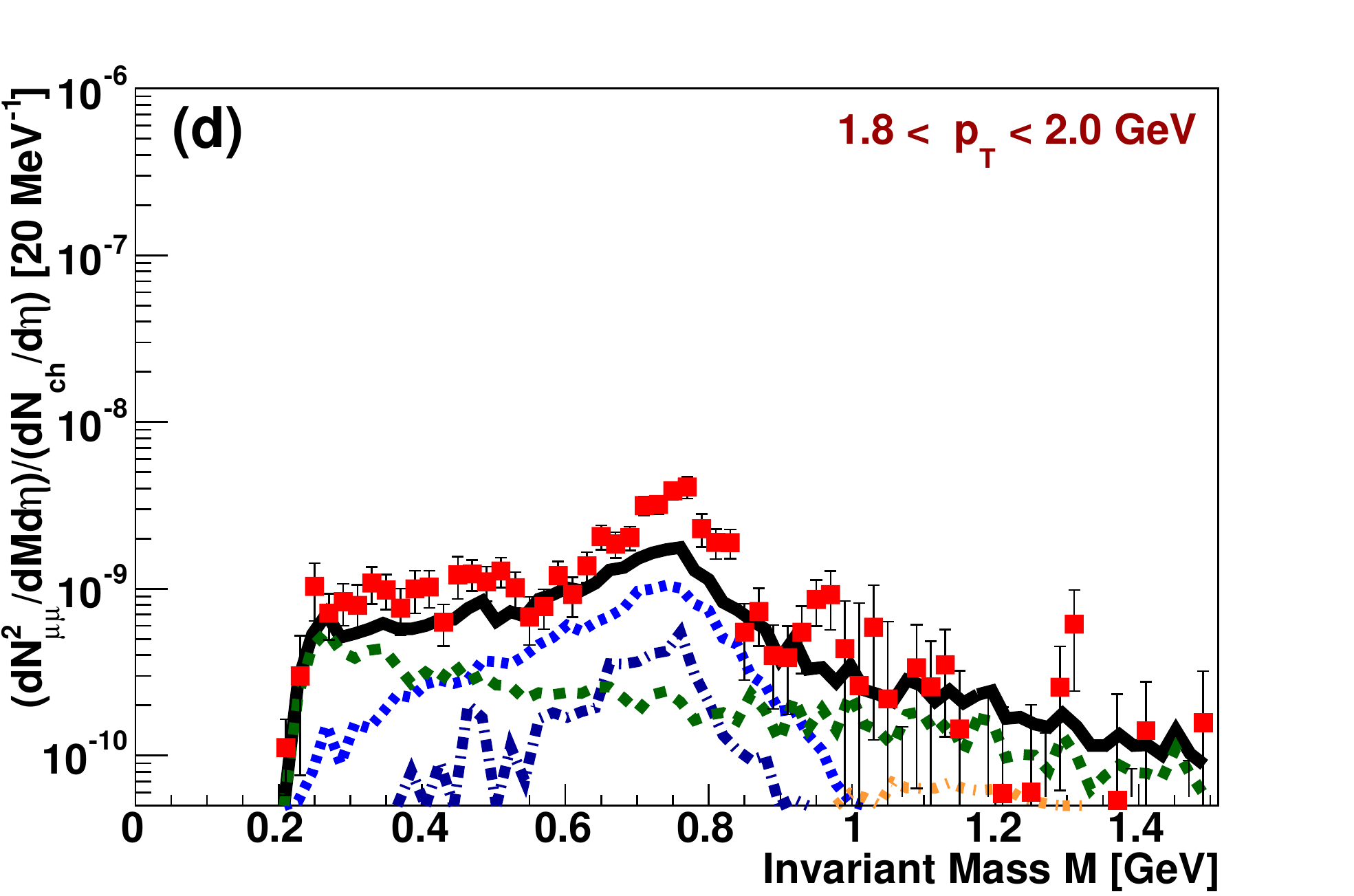}
\includegraphics[width=1.02\columnwidth]{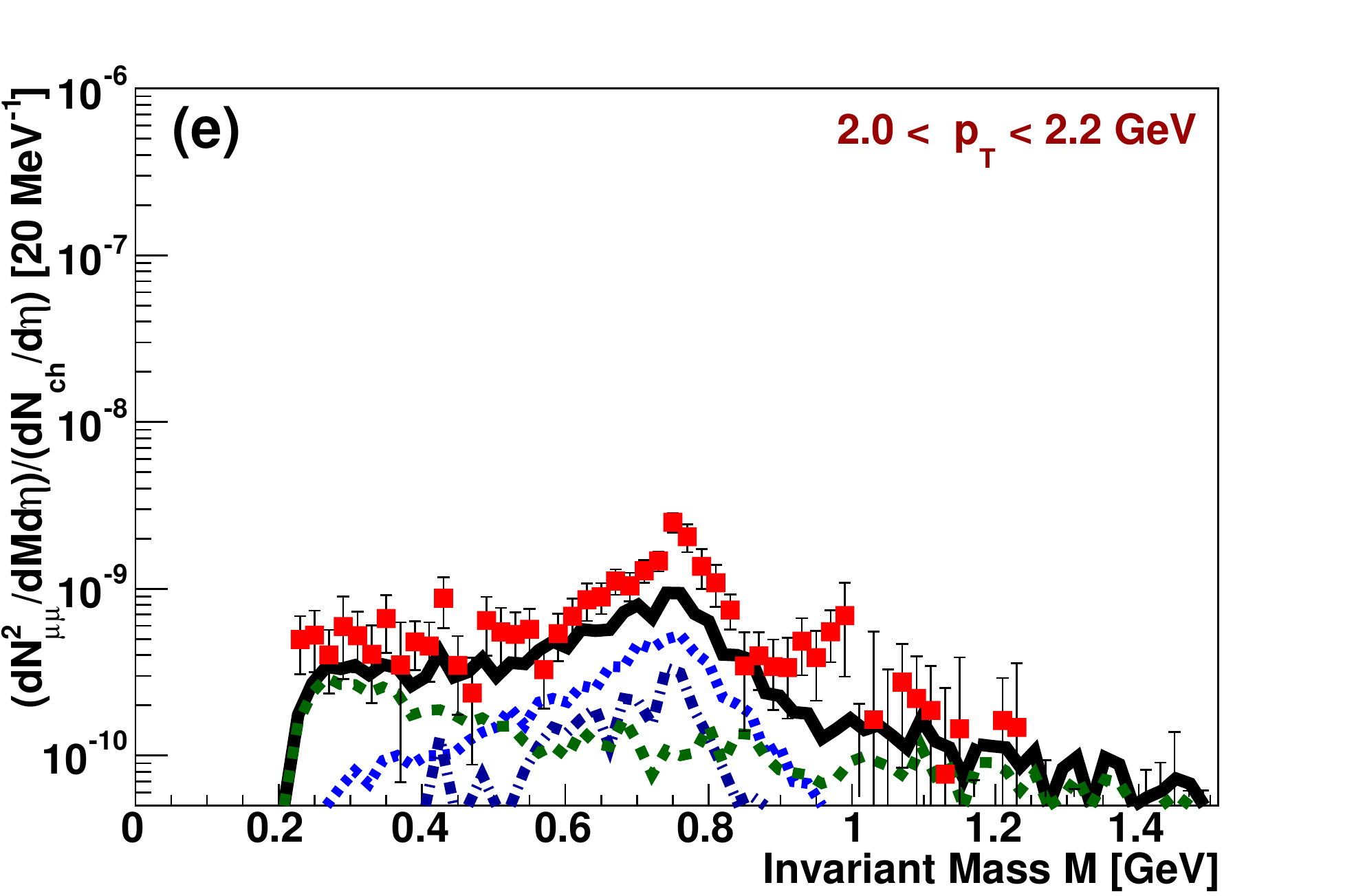}
\includegraphics[width=1.02\columnwidth]{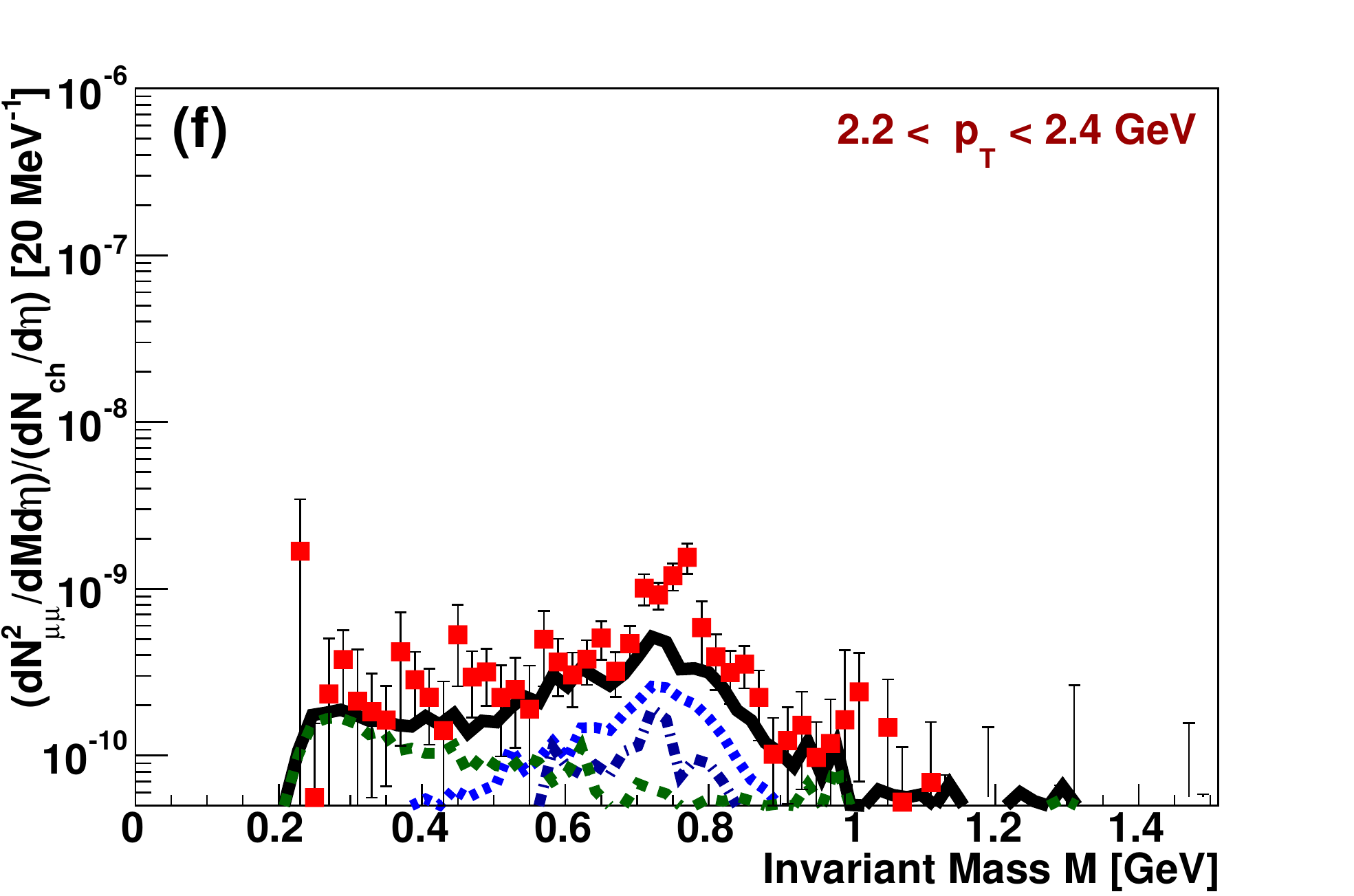}
\caption{\label{ptslices2} Same as in Figure \ref{ptslices1}, but for
  higher $p_{t}$-bins ranging from $p_{t}=1.2$ to $2.4\,\GeV$. The
  different contributions are the same as in Figures \ref{invmass} and
  \ref{mtspectra}. The results are compared to the experimental data
  from the NA60 Collaboration \cite{Arnaldi:2008fw}.}
\end{figure*}
As we can see from these profiles, the highest initial baryon and energy
densities should be expected at the origin of our grid in the center of
the collision. Indeed the time evolution of $\varepsilon$ and $\rho$ in
Figure \ref{timeev} (a) shows a rapid rise up to baryon and energy
densities of 20 times respectively 100 times the ground state
densities. The maxima are reached at a time of about $1.2\,\fm/c$ after
the beginning of the collision and in the following the densities
decrease again -- first rapidly, then slower again. After application of
the equation of state we observe a similar behavior for the time
dependence of the temperature, with values of up to $300\,\MeV$. It is
important to bear in mind that we apply two different EoS here, a
lattice EoS for temperatures above 170\,MeV to mimic a
Quark-Gluon-Plasma phase and a Hadron-Gas EoS for lower temperatures. As
we see, a smooth transition in temperature between the two EoS is obtained
around $t=5 \,\fm/c$, which one would have already expected from the
fact that the EoS agree quite well for $T < 200 \, \MeV$. The
baryo-chemical potential from the hadron gas EoS slowly rises from a
value of 200\,MeV at the transition time up to a maximum of 400\,MeV in
the course of the evolution. This increase is similar to the findings of
\cite{Huovinen:2002im} and caused by the stronger relative decline of
the energy density compared to the baryon density. A different behavior
can be observed for the pion chemical potential $\mu_{\pi}$, which is
around 100 MeV first and then drops to values around 0. This is unlike
the findings of fireball approaches, where particle numbers are fixed at
a freeze-out and the subsequent cooling of the system leads to a
build-up of a finite pion chemical potential. The picture in our
transport approach is completely different, as pions can be produced and
absorbed over the whole evolution in the system.

As we see from the time evolution of the central cell, the temperature
reaches values of 100\,MeV even after a time of $15\,\fm/c$. However,
this is a special case and for most cells the temperature has already
dropped beyond significance before. But in contrast to many approaches
with a fixed lifetime of the fireball, here an underlying microscopic
transport description is applied which takes into account that some
singular cells still reach quite high temperatures and densities even
after the usually assumed fireball lifetimes. The contribution to the
dilepton yield from these few cells is, however, quite negligible.

\subsection{\label{ssec:Results} Dilepton spectra}
The next step is to investigate how the space-time evolution obtained by
coarse-graining the transport simulations is reflected within the
resulting dilepton spectra. It is hereby of particular interest whether
and how the differences in the reaction dynamics as compared to the
fireball parametrizations will be reflected in the
$\mu^+\mu^-$-distributions as measurable in experiment (and whether one by
this can discriminate between different scenarios of the fireball
evolution).

In Figure \ref{invmass} the resulting dimuon invariant-mass spectra from
the coarse-graining calculations are compared to data from the NA60
Collaboration. There the dimuon excess yield in In+In collisions at a
beam energy of $E_{\text{lab}} = 158\,A \GeV$ with
$\erw{\dd N_{\mathrm{ch}}/\dd\eta}=120$ is shown, for the low-mass
region up to $M = 1.5$\,GeV (a) and the intermediate-mass regime up to
2.8\,GeV (b). We show the contributions of the in-medium $\rho$
emission, from the quark-gluon plasma, i.e., $q\bar{q}$-annihilation,
and the emission from multi-pion reactions, taking vector-axial-vector
mixing into account. The dilepton emission due to decays of $\rho$
mesons from the low-temperature cells is included as well, but in the
full $p_{t}$-integrated spectrum it is rather negligible compared to the
other contributions. Comparison with the experimental data from the NA60
Collaboration \cite{Arnaldi:2008fw, Specht:2010xu} shows a very good
agreement of our theoretical result with their measurement. Only a
slight tendency to underestimate the data in the invariant mass region from 
0.2 to 0.4 GeV and a minor excess above the data in the pole region is
observed. As the low-mass enhancement and the melting of the peak at the
pole mass are mainly caused by the baryonic effects on the $\rho$ meson
spectral function and very sensitive to the presence of baryons and
anti-baryons, this might be due to the fact that the baryon densities
(respectively the baryon chemical potential) are still slightly too low
in our approach. An additional modification of the spectral shape not
considered here may also be caused by the $\omega$-$t$-channel
exchange. It has been found, however, to give only a small contribution
to the total yield and is significant only for high transverse momenta
\cite{vanHees:2007th}. Furthermore, one has to bear in mind as well that
there is an uncertainty of up to 15\% around
$M \approx 0.4 \,\GeV$ between the parametrized spectral
function and its full evaluation from thermal field theory which has
been found in a full comparison between both approaches \cite{RappSF},
as mentioned above. Taking this and the systematic uncertainties of the
experimental data and of the model calculations into account, we
conclude that the approach is fully able to describe the total NA60
invariant mass spectrum with excellent accuracy.

To get an impression of the dominant role of baryon-induced medium
modifications, the thermal $\rho$ contribution assuming the absence of
all baryons and anti-baryons (i.e., for $\rho_{\text{eff}}=0$) is also
shown in Figure \ref{invmass}. In this case only meson-gas effects have
an influence on the spectral function. Compared to the full in-medium
$\rho$, it exhibits slightly more strength at the $\rho$ meson's pole
mass but is significantly below the experimental yield for
$M < 0.6$\,GeV by a factor of 2-5. Clearly, only the inclusion of
interactions with baryonic matter can explain the low-mass dilepton
excess, as has been noticed in previous studies \cite{vanHees:2007th}.

Comparing the dilepton emission rates obtained from the lattice and from
perturbative $q\bar{q}$-annihilation, both rates are identical for
masses larger than 0.8\,MeV, while the non-perturbative effects included
in the lattice calculations give rise to a strong increase (up to a
factor 3) of the yield at lower invariant masses. It is notable that the
shape of the slope in the region $M > 1.5\,\GeV$ is described with very
good accuracy.  This is important, as the hadronic contribution which
dominates at lower masses becomes negligible here, and the yield is
dominated by emission from the QGP phase. The intermediate-mass region
is therefore a good benchmark for a correct description of the
$q\bar{q}$ emission and allows for a reliable determination of the
space-time averaged temperature without distortion from blue-shift
effects due to flow (as is the case for effective slopes of $p_t$
spectra) \cite{Rapp:2014hha}. Note, however, that in contrast to the
results from a fireball approach we here get a slightly larger
contribution from the hadronic domain ($\rho$ and multi-$\pi$), whereas
in the fireball approach a more dominant QGP contribution is found,
especially at low masses \cite{vanHees:2007th, Rapp:2014hha}. This
finding strengthens the hypothesis of duality between the hadronic and
partonic dilepton emission in the transition temperature region between
both phases \cite{Rapp:1999ej}.

In Figure \ref{mtspectra} we present the transverse-mass ($m_{t}-M$)
spectra in four different mass bins of the dimuon-excess yield in In+In
collisions. The results are shown for mass bins of
$0.2 \, \GeV<M<0.4\,\GeV$ (a), $0.4 \, \GeV<M<0.6\,\GeV$ (b),
$0.6 \, \GeV<M<0.9\,\GeV$ (c), and $1.0 \,\GeV<M<1.4\,\GeV$ (d). The
different contributions are the same as in Figure
\ref{invmass}. Comparison to the NA60 results \cite{Damjanovic:2007qm}
shows again a very good agreement. The calculations are in almost all
cases within the error bars of the experimental data. Interesting is the
dominance of the different contributions in certain transverse-mass
respectively -momentum ranges. While the in-medium $\rho$ dominates the
spectra at low $m_{t}$ for all but the highest mass bin, the QGP and the
non-thermal $\rho$ do not significantly contribute at low $m_{t}$ but
their relative strength increases when going to higher transverse mass.

Besides the invariant-mass and the transverse-mass spectra, we also
studied the former as resolved in different $p_{t}$-slices. This
analysis is of special interest as theoretical studies show that the
medium modifications of the $\rho$ spectral function depend strongly on
the momentum. While a significant change of the spectral shape is
predicted based on constraints from vacuum scattering and decay data at
low momenta, this effect should become less and less significant for
higher momenta \cite{Leupold:2009kz}. The resulting invariant mass
spectra are shown in Figures \ref{ptslices1} and \ref{ptslices2}, with
12 different plots representing the different $p_{t}$-bins with a width
of $\Delta p_{t}=200\,\MeV$, ranging from the lowest values
$0.0\leq p_{t} \leq 0.2\,\GeV$ up to the highest transverse momentum bin
with $2.2 \, \GeV \leq p_{t} \leq 2.4 \,\GeV$. Our calculations once
again agree with the data and especially show the clear momentum dependence of
the $\rho$ contribution in the region below the meson's pole mass. While
for the lowest $p_{t}$-bin the yield in the mass range from 0.2\,GeV to
0.4\,GeV exceeds even the yield at the pole mass, this excess becomes
increasingly smaller when we go to higher transverse momenta. For the
highest $p_{t}$-bin the shape of the $\rho$ in the invariant mass
spectrum looks almost as in the vacuum. Besides, the relative
contribution from the non-thermal transport $\rho$ is increasing when
going to higher transverse momentum. It is in addition noteworthy that
also the non-thermal $\rho$, for which no explicit in-medium
modifications are implemented in UrQMD, shows dynamically some
$p_{t}$-dependence of the spectral shape. However, this is not
surprising since the transport model includes effects like resonance
excitation, rescattering or reabsorption that can cause such a momentum
dependent mass distribution, i.e., the spectral properties of the
transport $\rho$ include some medium effects and thus differ from those
in the vacuum. The very same microscopic mechanisms, of course, also
cause the medium modifications of the spectral functions within the
thermal quantum-field theoretical models.

Yet, for $p_{t}$ greater than 1.2\,GeV the yield in the pole mass region
of the $\rho$ meson, i.e.~at $M \approx 770 \, \MeV$, is still not
described fully. The experimental data show a more prominent and sharper
peak structure than we find within our approach. Some of the
"freeze-out" respectively "vacuum" $\rho$ contribution might be missing
in spite of including the non-thermal transport $\rho$.

In general it is interesting to see how the correct description of all
the three different thermal contributions is necessary to achieve
agreement with the data over the whole transverse momentum range. For
example, at the lowest masses (below 0.4\,GeV) the broadened $\rho$
delivers the significant contribution for low $p_{t}$, while at higher
$p_{t}$ the emission from the deconfined phase dominates at these
masses. This is another good benchmark that shows that we obviously
describe the thermal emission quite realistically. Nevertheless, one
also has to stress that the results for the total dilepton spectra
obtained in the present study agree with the studies performed with
fireball parametrizations, though the space-time evolution shows
significant differences between the two models. This indicates that the
time-integral nature of dilepton spectra to a large extent disguises the
details of the reaction dynamics by averaging over volume and lifetime.
\section{\label{sec:Summary} Conclusions \& outlook}

In this paper we have presented a coarse-graining approach to the
calculation of dilepton production in heavy-ion collisions. Using an
ensemble of several events from transport calculations with the UrQMD
model, we put the output on a space-time grid of small cells. By
averaging the particle distribution in each cell over a large number of
events and going into the local rest frame we can calculate the energy
and baryon density and consequently temperature and chemical potential
by introducing an equation of state. When the thermodynamic properties
of the cell are known the corresponding thermal dilepton emission rates
can be determined.  With this procedure it is aimed to achieve a more
realistic description of dilepton production in heavy-ion
collisions. Since a complete non-equilibrium treatment of medium
modifications is an extremely difficult task, the coarse-graining
approach is intended as a compromise to apply in-medium spectral
functions in combination with a microscopic description of the bulk
evolution of a heavy-ion collision.

The agreement between our results for thermal dilepton
invariant- and transverse-mass spectra and the experimental findings of
the NA60 collaboration is very good. The coarse-graining study
also confirms previous calculations with the same spectral function
within a fireball approach
\cite{vanHees:2006ng,vanHees:2007th,Rapp:2014hha}. However, it is
remarkable that in spite of differences in the dynamics of the
reaction, the final results are so similar in both approaches. The main
distinctions are, in summary:
\begin{enumerate}[(i)]
\item The rise of large chemical potentials in the earlier stages of the
  reaction within the coarse-graining approach, while in the fireball
  model a finite $\mu_{\pi}$ shows up after the freeze-out.
\item A larger fraction of QGP dilepton contribution is found in the
  fireball model while we get less QGP and more hadronic emission when
  coarse-graining the microscopic dynamics.
\item The lifetime of the hot and dense system is about 7\,fm/$c$ in the
  fireball parametrization while we still find thermal emission even
  after much longer time of 15\,fm/$c$ in the present study. Note hereby
  that for the latter case $T$, $\mu_{B}$ (respectively $\rho_{\mathrm{eff}}$) 
  and $\mu_{\pi}$ are determined locally whereas the fireball model 
  assumes global thermal equilibrium.
\end{enumerate}
The obvious explanation for the agreement is that the dilepton spectra
are only time-integrated results and therefore less significant with
regard to the very details of the reaction evolution but rather the
global scale of the dynamics, i.e., one is mainly sensitive to the
average thermal properties of the system. The high-mass tail of the
invariant mass spectrum, which is clearly dominated by the QGP emission
(i.e. for $M > 1.5$\,GeV) is a good example for this. It reflects the 
true average temperature of the source (without blue shifts as in the 
photon case) and was found to be roughly 205\,MeV in fireball
models. Looking at the details, one finds however that the QGP yield for
very high masses over 2.5\,MeV is larger in the coarse-graining, but the
overall slope of the partonic emission is flatter so that the 
differences between the two approaches show up especially at lower masses. 
This can be explained by the fact that we have some very hot cells with 
temperatures above 300 MeV populating the yield at very high masses, while 
the initial temperature in the fireball is only 245 MeV. In contrast, the 
overall QGP emitting volume is larger in the fireball model due to the
assumption of global equilibrium, while only a limited number of cells
reaches above $T_{c}$ in the coarse-graining approach. For lower masses,
less QGP yield is counter-balanced by a larger hadronic contribution. This
is not surprising, because the lower number of high-temperature cells
corresponds to a larger fraction of low-temperature emission. A more
detailed comparison beyond this will be addressed in a future
work. However, it becomes already clear that there are several
aspects which only show up in their combined effect in the dilepton
spectra, so e.g. a smaller volume can be compensated by a longer
lifetime of the fireball.

Nevertheless, in spite of the insensitivity with regard to the detailed
reaction dynamics, we can also draw some conclusions by the agreement of
the results from the fireball and coarse-graining model:
\begin{enumerate}
\item The large influence of baryons on the spectral shape of the $\rho$
  which is clearly responsible for the enhancement of the dilepton yield
  in the mass range $0.2\,\GeV < M_{\mu\mu} < 0.6\,\GeV$. The most
  significant modifications of the spectral function of the $\rho$ are
  found at low momenta, in line with previous experimental and
  theoretical investigations.
\item Thermal emission from the QGP, at least in parts for temperatures
  significantly above the critical temperature $T_{\mathrm{c}}$. Without
  this both models fail to fully explain the invariant mass spectrum at
  higher invariant masses.
\item Especially for the mass region above the $\rho$ meson, i.e, for
  $1$\,GeV$ < M_{\mu\mu} < 2$\,GeV the results also strengthen the
  hypothesis of quark-hadron duality, i.e., that the thermal emission
  rates are dual in the temperature range around the transition
  temperture between the partonic and the hadronic phase.
\end{enumerate}

For the future the latter two points deserve further investigation,
especially with regard to the transition from the hadronic to the
partonic phase. The present results show that it will be hard to
definitely determine details of the evolution of the reaction by
means of dilepton spectra.  But the question whether electromagnetic
probes can give hints for the creation of a QGP phase or whether
duality prohibits to discriminate between hadronic and partonic emission
in the transition region might be clarified in theoretical studies at
lower collision energies as, e.g., covered by the Beam-Energy Scan
program at the Relativistic Heavy Ion Collider (RHIC) and the future
FAIR facility (with $E_{\mathrm{lab}} = 8-35$\,AGeV). But also a full
understanding of the medium modifications of hadron properties and the
possible restoration of chiral symmetry has not yet been obtained and
further theoretical and experimental studies are desirable, especially
exploring the high-$\mu_{B}$ region of the QCD phase diagram.

Considering the aspects mentioned above, the fundamental applicability of
the coarse-graining approach for all kinds of collision energies opens
the possibility for a broad variety of future investigations. For
low-energy heavy-ion collisions as investigated at the GSI SIS (HADES)
it offers a unique option, since an application of conventional
hydrodynamic or fireball models seems not reasonable here and
microscopic transport models failed to give an unambiguous explanation
of the observed dilepton spectra at low bombarding energies. The
expected high baryon densities in these cases make a detailed study of
the thermodynamic properties interesting.

\begin{acknowledgments} 
  The authors especially thank Ralf Rapp for providing his spectral
  function and for many fruitful discussions. This work was supported by
  the Hessian Initiative for Excellence (LOEWE) through the Helmholtz
  International Center for FAIR (HIC for FAIR), the Federal Ministry of
  Education and Research (BMBF) and the Helmholtz Research School for
  Quark-Matter Studies (H-QM).
\end{acknowledgments}

\bibliography{Draft_NA60}

\end{document}